\documentclass[a4paper,11pt]{article}
\usepackage[a4paper,top=2.5cm,bottom=2.4cm,left=2.4cm,right=3cm]{geometry}
\pdfoutput=1 
\usepackage{jcappub}
\usepackage{bm}
\usepackage{graphicx}
\usepackage{multirow}
\usepackage{amsmath}
\usepackage{amssymb}
\usepackage{hyperref}
\usepackage{color}
\usepackage{xspace}
\usepackage{verbatim}
\usepackage{mathtools}
\usepackage{booktabs}
\usepackage[dvipsnames]{xcolor}
\usepackage[caption=false]{subfig}

\newlength{\dhatheight}

\newcommand{\D}[2][]{\ensuremath{\operatorname{d}^{#1}\!{#2}}}

\title{\huge Dark Matter EFT Landscape Probed by QUEST-DMC}
 \author[a]{\large QUEST-DMC Collaboration: N. Darvishi,}
 \author[b]{S. Autti,}
 \author[c]{L. Bloomfield,}
 \author[a]{A. Casey,}
 \author[a]{N. Eng,}
 \author[c]{P. Franchini,}
 \author[b]{R. P. Haley,}
 \author[a]{P. J. Heikkinen,}
 \author[e]{A. Jennings,}
 \author[f]{A. Kemp,}
 \author[c,a]{E. Leason,}
 \author[c]{J. March-Russell,}
 \author[b]{A. Mayer,}
 \author[c]{J. Monroe,}
 \author[b]{D. M{\"u}nstermann,}
 \author[b]{M. T. Noble,}
 \author[b]{J. R. Prance,}
 \author[a]{X. Rojas,}
 \author[b]{T. Salmon,}
 \author[a]{J. Saunders,}
 \author[d]{J. Smirnov,}
 \author[a,c]{R. Smith,}
 \author[b]{M. D. Thompson,}
 \author[b]{A. Thomson,}
 \author[a]{A. Ting,}
 \author[b]{V. Tsepelin,}
 \author[a]{S. M. West,}
 \author[b]{L. Whitehead,}
 \author[b]{D. E. Zmeev}

 \affiliation[a]{\textit{Department of Physics, Royal Holloway University of London, Egham, Surrey, TW20 0EX, UK}}
\affiliation[b]{\textit{Department of Physics, Lancaster University, Lancaster, LA1 4YB, UK}}
\affiliation[c]{\textit{Department of Physics, University of Oxford, Keble Road, Oxford, OX1 3RH, UK}}
\affiliation[d]{\textit{Department of Mathematical Sciences, University of Liverpool, Liverpool, L69 7ZL, UK}}
\affiliation[e]{\textit{RIKEN Center for Quantum Computing, RIKEN, Wako, 351-0198, Japan}}
\affiliation[f]{\textit{UKRI STFC Rutherford Appleton Laboratory, Particle Physics Department, Harwell, Didcot OX11 0QX, UK}}

\emailAdd{neda.darvishi@rhul.ac.uk}

\bigskip\abstract{\noindent
\noindent We present the projected sensitivity to non-relativistic Effective Field Theory (EFT) operators for Dark Matter (DM) direct detection using the QUEST-DMC experiment. QUEST-DMC employs superfluid Helium-3 as a target medium and measures energy deposition via nanomechanical resonators with SQUID-based readout to probe DM interactions. The experiment aims to explore new parameter space in the sub-GeV mass range, probing light DM and a broad range of interaction models. We analyse the sensitivity to a complete set of fourteen independent non-relativistic EFT operators, each parameterised by a Wilson coefficient that quantifies the strength of DM interactions with Standard Model particles. For each interaction channel, we determine the corresponding sensitivity ceiling due to attenuation of the DM flux incident on the detector, caused by DM scattering in the Earth and atmosphere. As a key component of this analysis, we provide the mapping between the non-relativistic EFT operators and the relativistic bilinear DM–nucleon interactions, and assess the interaction sensitivity to sub-GeV DM in the QUEST-DMC detector.
Our findings demonstrate that QUEST-DMC provides a unique probe of DM interactions, particularly in previously unexplored parameter space for momentum- and velocity-dependent interactions, thereby expanding the search for viable DM candidates beyond traditional weakly interacting massive particles. 
}

\begin{document}
\maketitle
\flushbottom

\section{Introduction}\label{sec:introduction}

Dark Matter (DM) is a significant component of the universe and plays a fundamental role in cosmic evolution. Its gravitational influence has been essential in forming large-scale structures and maintaining galaxies' stability.

Traditional direct detection experiments have primarily focused on Weakly Interacting Massive Particles (WIMPs) with masses above that of the proton. Despite extensive efforts, direct detection of DM interactions in terrestial experiments remains elusive. This has led to increasing interest in lighter DM candidates, particularly those with masses around or below the proton mass. Various well-motivated theories~\cite{Zurek:2013wia,Barnes:2020vsc,Hochberg:2014dra,Pospelov:2008jk,Jaeckel:2012mjv,Hall:2009bx} support this shift, driving direct detection searches for sub-GeV DM scattering.
This requires experiments with exceptionally low energy thresholds, below the keV scale. As the recoil energy in DM-nucleus interactions is inversely proportional to the target nucleus mass, low-mass nuclei are advantageous for detection.

The earlier work~\cite{QUEST-DMC:2023nug,Autti:2023gxg,Autti:2024awr} introduced the fundamental principles of the Quantum Enhanced Superfluid Technologies for Dark Matter and Cosmology (QUEST-DMC) experiment that is searching for sub-GeV DM. The experiment operates with a superfluid He-3 target at temperatures below 200 microKelvin, and employs nanomechanical resonators for detection, read out by SQUIDs. The relatively light He-3 target facilitates both Spin-Independent (SI) and Spin-Dependent (SD) interaction searches for DM, which allows for a comprehensive exploration of DM candidates, particularly in the sub-GeV mass range. 

QUEST-DMC is designed to achieve world-leading sensitivity to small SD DM-neutron scattering cross-sections and to SI DM-nucleon interactions for masses below $0.025~{\rm GeV}/c^2$. In Ref.~\cite{QUEST-DMC:2025qsa}, we extended this analysis to scenarios with large DM interaction cross sections, exploring the experiment's sensitivity limits and benchmarking its performance against competing detection techniques. Here, we examine how these sensitivities vary across different interaction types, parameterised within a generalised non-relativistic Effective Field Theory (EFT) approach, following the theoretical frameworks of Refs.~\cite{Fan:2010gt,Fitzpatrick:2012ix}, and using the conventions described in Ref.~\cite{Anand:2013yka}. In the context of non-relativistic EFT for DM, Wilson coefficients play a crucial role in describing the interactions between DM and standard model particles.
The interaction Hamiltonian is expressed as
\begin{align}
\hat{{\cal H}}=\sum_{\tau=0,1}\sum_{i=1}^{15} c_i^\tau {\cal O}_i t^\tau,
\end{align}
where $c_i^\tau$ are the Wilson coefficients and ${\cal O}_i$ represent the corresponding operators. The isospin operators $t^0$ and $t^1$ correspond to $\sigma^0$ and $\sigma^3$, respectively, of the Pauli matrices.

This approach is more general than the standard SI and SD couplings typically reported by experiments in the literature, incorporating momentum-dependent and velocity-dependent operators. Within the framework of elastic DM He-3 interactions, which adhere to momentum conservation and Galilean invariance, the operators can be distilled into six fundamental Hermitian quantities:
$$
\vec{1}_{\chi}, \quad \vec{1}_{N}, \quad i \frac{\vec{q}}{m_N}, \quad \vec{v}^\perp , \quad \vec{S}_{\chi}, \quad \vec{S}_{N},
$$
where $\vec{1}_{{\chi}}$ and $\vec{1}_{N}$ are identity operators acting on the DM and the target respectively. The transverse relative velocity of the incoming DM particle and the target is defined as $\vec{v}^{\perp} \equiv \vec{v} + \vec{q}/2\mu_{\chi N}$, where $\vec{v}$ is the DM velocity in the lab frame, $\vec{q}$ is the momentum transfer, $m_N$ is the nuclear mass and $\mu_{\chi N} = m_{\chi} m_N/(m_{\chi} + m_N)$ is the DM-nucleus reduced mass. $\vec{S}_{\chi}$ and $\vec{S}_N$ denote the spins of the DM particle and the nucleus, respectively.

Combining these elements up to second order in $\vec{q}$, we obtain fifteen independent and dimensionless EFT operators. However, the operator $\mathcal{O}_2$ is omitted because it cannot arise in the non-relativistic limit from a relativistic operator to leading order. Thus, the fourteen operators are listed as ~\cite{Fan:2010gt,Fitzpatrick:2012ix}:
\begin{equation}
\begin{array}{l}
\mathcal{O}_{1}=1_\chi 1_N, \\
\mathcal{O}_{3}=i \vec{S}_{N} \cdot(\frac{\vec{q}}{m_N} \times \vec{v}^{\perp}), \\
\mathcal{O}_{4}=\vec{S}_{\chi} \cdot \vec{S}_{N}, \\
\mathcal{O}_{5}=i \vec{S}_{\chi} \cdot(\frac{\vec{q}}{m_N} \times \vec{v}^{\perp}), \\
\mathcal{O}_{6}=(\vec{S}_{\chi} \cdot \frac{\vec{q}}{m_N})(\vec{S}_{N} \cdot \frac{\vec{q}}{m_N}), \\
\mathcal{O}_{7}=\vec{S}_{N} \cdot \vec{v}^{\perp}, \\
\mathcal{O}_{8}=\vec{S}_{\chi} \cdot \vec{v}^{\perp}, \\
\mathcal{O}_{9}=i \vec{S}_{\chi} \cdot(\vec{S}_{N} \times \frac{\vec{q}}{m_N}), \\
\mathcal{O}_{10}=i \vec{S}_{N} \cdot \frac{\vec{q}}{m_N}, \\
\mathcal{O}_{11}=i \vec{S}_{\chi} \cdot \frac{\vec{q}}{m_N}, \\
\mathcal{O}_{12} =\vec{S}_{\chi} \cdot(\vec{S}_{N} \times \vec{v}^{\perp}), \\
\mathcal{O}_{13} =i(\vec{S}_{\chi} \cdot \vec{v}^{\perp})(\vec{S}_{N} \cdot \frac{\vec{q}}{m_N}), \\
\mathcal{O}_{14} =i(\vec{S}_{\chi} \cdot \frac{\vec{q}}{m_N})(\vec{S}_{N} \cdot \vec{v}^{\perp}), \\
\mathcal{O}_{15} =-(\vec{S}_{\chi} \cdot \frac{\vec{q}}{m_N})\left((\vec{S}_{N} \times \vec{v}^{\perp}) \cdot \frac{\vec{q}}{m_N}\right).~~~~~~~~~~~
\end{array}
\label{operators}
\end{equation}
Each of these operators can in principle couple differently to protons versus neutrons (or equivalently, to isoscalars versus isovectors); however, for He-3 in the case of SD interactions $\langle S_p \rangle$ = 0, leaving only DM-neutron scattering.

\section{Modelling}\label{sec:modeling}
Using the profile likelihood ratio (PLR) framework developed in Ref.~\cite{QUEST-DMC:2023nug}, we use statistical inference to quantify the sensitivity of QUEST-DMC to identify or constrain the possibility of DM interacting under a given EFT operator. A likelihood ratio test requires a good model of both the null (background-only) and alternative (background plus DM interaction signal) to be valid.

The differential event rate with respect to recoil energy, $E_R$, induced by DM scattering in the detector, can be written in the following form:
\begin{align}
\label{RateFromMatrixElement}
\frac{\D R}{\D E_R} = \frac{\rho_\chi}{2 \pi m_\chi } \int_{v>v_{\rm min}} \frac{f(\vec{v})}{v} \overline{|\mathcal{M}(v^2, q^2)|^2} \D[3]{v},
\end{align}
where $m_\chi$ is the DM mass, $\rho_\chi = 0.3\,\rm GeV/c^2$ is the local DM density, and $f(\vec{v})$ is the DM velocity distribution in the detector rest frame. The lower limit of the integral is set by the minimum DM speed required to produce a recoil of energy $E_R$, given by $v_{\rm min} = ( m_N E_R / 2 \mu_{\chi N}^2)^{1/2}$. The velocity distribution $f(\vec{v})$ is taken to follow a truncated Maxwell-Boltzmann profile, consistent with the Standard Halo Model~\cite{Lewin:1995rx,Savage:2006qr}.

In Eq.~\eqref{RateFromMatrixElement}, $\overline{|\mathcal{M}|^2}$ represents the spin-averaged squared matrix element for a given interaction. 
It is obtained by averaging the squared amplitude $|\mathcal{M}|^2$ over the initial spins and summing over the final spins of the DM particle and the nucleon, according to
\begin{equation}
\overline{|\mathcal{M}(v^2, q^2)|^2} = {1 \over 2j_\chi + 1} {1 \over 2j_N + 1} \sum_\mathrm{spins} |\mathcal{M}|^2.
\label{eq:transitionprob}
\end{equation}
This expression can be equivalently rewritten in terms of the coefficients $c_i^\tau$ of non-relativistic EFT operator $i$ and nuclear response functions, where the velocity- and momentum-transfer dependence is captured by form factors $R_i^{\tau \tau^\prime}(\vec{v}_T^{\perp\,2}, \vec{q}^{\,2} / m_N^2)$ as
\begin{align}
\overline{|\mathcal{M}(v^2, q^2)|^2} &\equiv {4 \pi \over 2j_N+1} \times
\nonumber \\
& 
\hspace{-0.7cm}\sum_{k} \sum_{{\tau,\tau}^\prime=0,1} R^{\tau \tau^\prime}_k \left( \vec{v}_T^{\perp 2}, {\vec{q}^{\,2} \over m_N^2},\left\{c_i^\tau c_j^{\tau^\prime} \right\} \right)
~S_k^{\tau \tau^\prime}(y).
\label{Ptot}
\end{align}
The quantities $j_N$ and $j_\chi$ are the spins of the nucleus and the DM particle, respectively. Conventionally, $\tau, \tau' = 0, 1$ label the isospin indices, with $\tau = 0$ pertinent to the isoscalar and $\tau = 1$ to the isovector components.

The index $k$ labels nuclear response types that characterise how a nucleus responds to DM interactions~\cite{Fitzpatrick:2012ix}. For the He-3 nucleus, which has spin-$1/2$ and a relatively simple structure, the relevant contributions come from the $M$ response, associated with SI scattering, and from the transverse and the longitudinal SD responses $\Sigma'$ and $\Sigma^{\prime\prime}$.

Accordingly, the following DM response functions $R_k^{\tau\tau'}$ are relevant for scattering with He-3:
\begin{subequations}
\label{Eq:RF}
\begin{align}
 R_{M}^{\tau \tau^\prime} = & \, c_1^\tau c_1^{\tau^\prime } + \dfrac{j_\chi (j_\chi+1)}{3} \left[ \frac{q^2}{m_N^2} v_T^{\perp 2} c_5^\tau c_5^{\tau^\prime } + v_T^{\perp 2}c_8^\tau c_8^{\tau^\prime } + \dfrac{q^2}{m_N^2} c_{11}^\tau c_{11}^{\tau^\prime } \right], \\
 R_{\Sigma^{\prime \prime}}^{\tau \tau^\prime} = & \,\dfrac{j_\chi (j_\chi+1)}{12} \left[ c_4^\tau c_4^{\tau^\prime} + \dfrac{q^2}{m_N^2} (c_4^\tau c_6^{\tau^\prime }
 +c_6^\tau c_4^{\tau^\prime }) + \dfrac{q^4}{m_N^4} c_{6}^\tau c_{6}^{\tau^\prime }
 +\dfrac{q^2}{4 m_N^2} c_{10}^\tau c_{10}^{\tau^\prime } 
 \right. \\
 & \left. 
 +v_T^{\perp 2} c_{12}^\tau c_{12}^{\tau^\prime } 
 +\dfrac{q^2}{m_N^2} v_T^{\perp 2} c_{13}^\tau c_{13}^{\tau^\prime } \right], \nonumber \\
 R_{\Sigma^\prime}^{\tau \tau^\prime} = \, \dfrac{1}{8} & \left[ \dfrac{q^2}{m_N^2} v_T^{\perp 2} c_{3}^\tau c_{3}^{\tau^\prime } + v_T^{\perp 2} c_{7}^\tau c_{7}^{\tau^\prime } \right] + \dfrac{j_\chi (j_\chi+1)}{12} \,\bigg[ c_4^\tau c_4^{\tau^\prime} + \dfrac{q^2}{m_N^2} c_9^\tau c_9^{\tau^\prime } + \frac{q^2}{2 m_N^2} v_T^{\perp 2} c_{14}^\tau c_{14}^{\tau^\prime } \nonumber \\
 & +\dfrac{v_T^{\perp 2}}{2} \left(c_{12}^\tau-\dfrac{q^2}{m_N^2}c_{15}^\tau \right)\left( c_{12}^{\tau^\prime } -\dfrac{q^2}{m_N^2}c_{15}^{\tau \prime} \right) \bigg].
\end{align}
\end{subequations}
In Eq.~\eqref{Ptot}, the nuclear response functions for He-3~\cite{Catena:2015uha} take the following form
\begin{align}
S^{\tau\tau'}_k(y) = A^{\tau\tau'}_k \, e^{-2y},
\end{align}
with the non-zero coefficients, as
\begin{align}
\label{he3-NRF}
A^{00}_{\Sigma^\prime} &= A^{11}_{\Sigma^\prime}= -A^{01}_{\Sigma^{\prime}} = -A^{10}_{\Sigma^{\prime}} =0.0795775, \nonumber\\
A^{00}_{\Sigma^{\prime\prime}}&= A^{11}_{\Sigma^{\prime\prime}}=-A^{01}_{\Sigma^{\prime\prime}} = -A^{10}_{\Sigma^{\prime\prime}}=A^{11}_{M} =0.0397887, \nonumber\\
A^{00}_{M} &= 0.358099,\nonumber\\
A^{01}_{M} &=A^{10}_{M}= 0.119366, 
\end{align}
and $y=(b q)^2/4$ is a dimensionless variable, with length parameter $b=(8.2934/(9 A^{-1/3}-5A^{-2/3}))^{1/2}{~~\rm fm}$ and $A=3$.

The interaction can be rewritten in terms of contributions from proton ($p$) and neutron ($n$) instead of using the isoscalar and isovector components, with the relations $c_i^p = \tfrac{1}{2}(c_i^0 + c_i^1)$ and $c_i^n = \tfrac{1}{2}(c_i^0 - c_i^1)$. This decomposition can be similarly applied to the DM and nuclear response functions. In the general case, the DM response functions relevant to $pp$, $nn$, and $np$ interactions can be expressed in terms of the isospin components as
\begin{align}
R^{pp/nn} &= \dfrac{1}{4} \left( R^{00} + R^{11} \pm R^{01} \pm R^{10} \right), \qquad
R^{np}= \dfrac{1}{4} \left( R^{00} - R^{11} \right),
\end{align}
with the upper signs corresponding to $pp$ and the lower signs to $nn$, and with a similar structure for $S_k^{\tau \tau^\prime}(y)$:
\begin{align}
S^{pp/nn} &= S^{00} + S^{11} \pm S^{01} \pm S^{10}, \qquad
S^{np}= S^{00} - S^{11}.
\end{align}
Therefore, $\overline{|\mathcal{M}|^2}$ decompose into $pp$, $nn$ and $np$ in a generic form as 
\begin{align}
\overline{|\mathcal{M}(v^2, q^2)|^2} \equiv {4 \pi \over 2j_N+1} \sum_{k} R_k^{pp}S_k^{pp}+R_k^{nn}S_k^{nn}+2R_k^{np}S_k^{np}.
\end{align}
Substituting the values specific to He-3 for SD interactions from Eq.~\eqref{he3-NRF} shows that the $pp$ and $np$ structures vanish, leaving only the $nn$ response to be non-zero, consistent with the nuclear spin being primarily carried by the unpaired neutron.

While destructive interference between multiple non-relativistic operators can occur, we begin by following the standard convention of assuming that one coupling dominates. In this single-operator framework, all but one of the coefficients $c_i^{n,p}\to 0$, with $n, p$ for the neutron or proton. These single-operator projections are not intended to represent UV-complete theories, but rather serve as a clear benchmark for interpreting the detector response to different interaction types.

At low energies, DM interactions with Standard Model fields can be described either via a relativistic EFT involving couplings to quarks, gluons, and photons, or through a Galilean-invariant non-relativistic EFT in terms of nucleons. In a UV-complete framework, each relativistic operator typically matches onto a combination of non-relativistic operators rather than a single term. In Section~\ref{subsec:rel-couplings}, we follow the conventions of Ref.~\cite{Anand:2013yka} to detail the matching between relativistic EFT operators and the corresponding set of non-relativistic operators. For a more systematic connection to UV physics, one may consider a quark/gluon-level EFT framework (see e.g. Ref.~\cite{Bishara:2017pfq}), which goes beyond the nucleon-level approach adopted in this work.

Owing to the linear dependence of the differential rate on $(c_i^n)^2$, we recast the non-relativistic couplings in terms of point-like DM–neutron cross sections.
For each single operator $\mathcal{O}_i$ we define $\sigma^{c_i}_{\chi n}(v)$ as the total $\chi n$ cross section in the nucleon (point-like) limit at fixed lab speed $v$, with an angular average appropriate to isotropic elastic scattering and with the nuclear response frozen at $q\!\to\!0$.
In this convention, $\sigma^{c_1}_{\chi n}$ and $\sigma^{c_4}_{\chi n}$ are the usual velocity/momentum-independent SI and SD cross sections, while for $i\neq 1,4$ the cross sections are explicitly velocity- and/or momentum–dependent.
With these definitions, the normalisations read:
\begingroup
\setlength{\parindent}{0pt}
\par\noindent

{\allowdisplaybreaks
\begin{subequations}
\label{eq:ci-two-col}

\noindent\makebox[\textwidth][l]{%
\begin{minipage}[t]{0.52\textwidth}\raggedright
\small\setlength{\parindent}{-1in}
\begin{align}
(c_1^n)^2 &= \pi\,\frac{\sigma_{\chi n}^{c_1}}{\mu_{\chi n}^2}, \\
(c_3^n)^2 &= \frac{4}{S_{\Sigma'}(0)}\,\frac{\sigma_{\chi n}^{c_3}(v)}{\mu_{\chi n}^2}\,
\frac{1}{\langle q^2/m_N^2\, v_T^{\perp 2} \rangle}, \\
(c_4^n)^2 &= \frac{4\pi}{j_\chi (j_\chi+1)}\,\frac{\sigma_{\chi n}^{c_4}}{\mu_{\chi n}^2}, \\
(c_5^n)^2 &= \frac{3\pi}{j_\chi (j_\chi+1)}\,\frac{\sigma_{\chi n}^{c_5}(v)}{\mu_{\chi n}^2}\,
\frac{1}{\langle q^2/m_N^2\, v_T^{\perp 2} \rangle}, \\
(c_6^n)^2 &= \frac{6}{j_\chi (j_\chi+1)}\,\frac{1}{S_{\Sigma''}(0)}\,
\frac{\sigma_{\chi n}^{c_6}(v)}{\mu_{\chi n}^2}\,
\frac{1}{\langle q^4/m_N^4 \rangle}, \\
(c_7^n)^2 &= \frac{4}{S_{\Sigma'}(0)}\,\frac{\sigma_{\chi n}^{c_7}(v)}{\mu_{\chi n}^2}\,
\frac{1}{\langle v_T^{\perp 2} \rangle}, \\
(c_8^n)^2 &= \frac{3\pi}{j_\chi (j_\chi+1)}\,\frac{\sigma_{\chi n}^{c_8}(v)}{\mu_{\chi n}^2}\,
\frac{1}{\langle v_T^{\perp 2} \rangle}, \\
(c_9^n)^2 &= \frac{6}{j_\chi (j_\chi+1)}\,\frac{1}{S_{\Sigma'}(0)}\,
\frac{\sigma_{\chi n}^{c_9}(v)}{\mu_{\chi n}^2}\,
\frac{1}{\langle q^2/m_N^2 \rangle},
\end{align}
\end{minipage}\hfill
\begin{minipage}[t]{0.485\textwidth}\raggedright
\small\setlength{\parindent}{0pt}
\begin{align}
(c_{10}^n)^2 &= \frac{2}{S_{\Sigma''}(0)}\,\frac{\sigma_{\chi n}^{c_{10}}(v)}{\mu_{\chi n}^2}\,
\frac{1}{\langle q^2/m_N^2 \rangle}, \\
(c_{11}^n)^2 &= \frac{3\pi}{j_\chi (j_\chi+1)}\,\frac{\sigma_{\chi n}^{c_{11}}(v)}{\mu_{\chi n}^2}\,
\frac{1}{\langle q^2/m_N^2 \rangle}, \\
(c_{12}^n)^2 &= \frac{12}{j_\chi (j_\chi+1)}\,\frac{\sigma_{\chi n}^{c_{12}}(v)}{\mu_{\chi n}^2}\,
\frac{\pi}{3-\pi S_{\Sigma'}(0)}\,
\frac{1}{\langle v_T^{\perp 2} \rangle}, \\
(c_{13}^n)^2 &= \frac{6}{j_\chi (j_\chi+1)}\,\frac{1}{S_{\Sigma''}(0)}\,
\frac{\sigma_{\chi n}^{c_{13}}(v)}{\mu_{\chi n}^2}\,
\frac{1}{\langle q^2/m_N^2\, v_T^{\perp 2} \rangle}, \\
(c_{14}^n)^2 &= \frac{12}{j_\chi (j_\chi+1)}\,\frac{1}{S_{\Sigma'}(0)}\,
\frac{\sigma_{\chi n}^{c_{14}}(v)}{\mu_{\chi n}^2}\,
\frac{1}{\langle q^2/m_N^2\, v_T^{\perp 2} \rangle}, \\
(c_{15}^n)^2 &= \frac{12}{j_\chi (j_\chi+1)}\,\frac{1}{S_{\Sigma'}(0)}\,
\frac{\sigma_{\chi n}^{c_{15}}(v)}{\mu_{\chi n}^2}\,
\frac{1}{\langle q^4/m_N^4\, v_T^{\perp 2} \rangle}.
\end{align}
\end{minipage}%
}
\end{subequations}
}
\endgroup

Here $\sigma_{\chi n}^{\,1,5,8,11}$ are SI total cross sections and all others are SD.
Angle brackets $\langle\cdots\rangle$ denote fixed-speed angular averages for isotropic elastic scattering (e.g. $\langle v_T^{\perp 2}\rangle=v^2/2$, $\langle q^2/m_N^2\rangle=2(\mu_{\chi n}/m_N)^2 v^2$, etc.).
Except for $\mathcal{O}_1$ and $\mathcal{O}_4$, the point-like $\chi n$ cross sections scale with even powers of the speed after this average, i.e.
$\sigma^{c_i}_{\chi n}(v)\propto v^{2}$ for $\mathcal{O}_{7,8,9,10,11,12}$,
$\propto v^{4}$ for $\mathcal{O}_{3,5,6,13,14}$, and
$\propto v^{6}$ for $\mathcal{O}_{15}$. The predicted event rate follows Eq.~\eqref{RateFromMatrixElement}, i.e. we fold $\sigma^{c_i}_{\chi n}(v)$ with the halo speed distribution and the nuclear and DM response functions $S_k(y)$ and $R_k$ from Eq.~\eqref{Eq:RF}. 
For $\mathcal{O}_{12}$ the non-relativistic reduction yields a specific linear combination of spin responses, $\Sigma'$ and $\Sigma''$, which gives the overall factor $\pi/(3-\pi S_{\Sigma'}(0))$; the interaction remains purely SD with no interference with the SI response.
For SI scattering, we assume isoscalar couplings, $c_i^p = c_i^n$. In the SD case, only neutron couplings $c_i^n$ are considered.

\section{Attenuation of the Dark Matter Flux}
\label{sec:atmospheric_attenuation}

The QUEST-DMC detector is located at the surface, thus in addition to earth shadowing we consider atmospheric attenuation of the DM flux incident on the detector as well. This phenomenon becomes particularly relevant for large cross-section DM interactions, where multiple scattering events modify the velocity distribution and reduce the number of detectable DM particles~\cite{Collar:1992qc,Collar:1993ss,Hasenbalg:1997hs,Kouvaris:2014lpa,Kouvaris:2015laa,Bernabei:2015nia}. A precise understanding of these effects is essential for accurate sensitivity predictions in surface-based detection experiments.

The attenuation effect arises from the scattering of DM particles as they traverse the Earth and its atmosphere. At sufficiently large cross-sections, the Earth becomes effectively opaque to DM, halting particles arriving from below, so that the detectable flux is predominantly incident from above after passing only through the atmosphere.
The primary atmospheric constituents affecting DM attenuation are Nitrogen-14 (N-14) and Oxygen-16 (O-16), which together make up over 99\% of the atmosphere by volume. For SI interactions, both N-14 and O-16 contribute, while for SD interactions, N-14 is relevant.

To model DM interactions in the atmosphere, we adopt an exponential density profile: 
\addtocounter{equation}{+1}\begin{equation} n_{\rm N}(\mathbf{r}) = n_0~ \exp\left(-\frac{\mathbf{r} - R_E}{H}\right), \end{equation} where $n_0$ is the sea-level number density, $R_E \approx 6371$ km is the radius of the Earth, and the height of the atmosphere $H \approx 80$ km. This model provides a realistic description of how atmospheric density decreases with altitude, thereby influencing the probability of DM scattering before reaching a detector.

The treatment of DM attenuation here assumes the straight-line path (SLP) approximation, in which DM particles are assumed to travel directly through the atmosphere, continuously losing energy, while never deviating from the SLP. This assumption is computationally efficient and provides reasonable estimates for heavier DM particles, which undergo minimal deflection due to their large mass-to-target mass ratio~\cite{Kouvaris:2015laa,Kavanagh:2017cru}.
For sub-GeV DM, a more accurate description is obtained using a diffusion framework which accounts for the effects of multiple random scatterings. In this description, DM particles experience both energy loss and angular diffusion, modifying the velocity spectrum at the detector~\cite{Cappiello:2023hza}. Although the diffusion framework offers a more rigorous treatment, the difference in projected sensitivity limits compared to the SLP approximation is approximately a factor of two. For a detailed analysis, see Ref.~\cite{QUEST-DMC:2025qsa}, where we show the ceiling for SI and SD interaction corresponding to the operators $\mathcal{O}_{1}$ and $\mathcal{O}_{4}$, respectively, calculating in both the SLP and diffusive frameworks. The SLP approximation provides a sufficiently accurate sensitivity ceiling for our purpose here, though we note that diffusion effects become important when precision constraints are required, particularly at low masses. 

In addition to the $M$, $\Sigma^{\prime}$, and $\Sigma^{\prime\prime}$ responses relevant for He-3, the atmospheric attenuation calculation includes additional responses arising from the more complex nuclear structure of atmospheric elements for some operators. These include the spin-orbit response $\Phi^{\prime\prime}$, the angular-momentum–dependent term $\tilde{\Phi}^\prime$, the orbital current response $\Delta$, and interference terms such as $M\Phi^{\prime\prime}$ and $\Delta\Sigma^\prime$. The corresponding DM and nuclear response functions, $R_k^{\tau\tau'}$ and $S_k^{\tau\tau'}$, are incorporated for N-14 and O-16, allowing for a consistent evaluation of the interaction cross section associated with each EFT operator, across atmospheric shadowing as well as scattering in the detector.

\section{Results}\label{sec:results}
To project sensitivities for various interaction coefficients, we incorporate the QUEST-DMC detector response model described in Ref.~\cite{QUEST-DMC:2023nug} into the expected event rate for each operator. This includes uncertainties from readout noise, fluctuations in quasiparticle production, and shot noise, all bounded by the detector's threshold energy. As in Ref.~\cite{QUEST-DMC:2023nug}, an energy threshold of 31 eV is assumed for the conventional readout with a cold transformer, and 0.51 eV for the SQUID-based readout. 
The assumed exposure here is $4.9$~g$\cdot$day corresponds to five 0.03 g He-3 cells operated over a six-month cycle.

\begin{figure*}[t]
\includegraphics[width=0.495\textwidth]{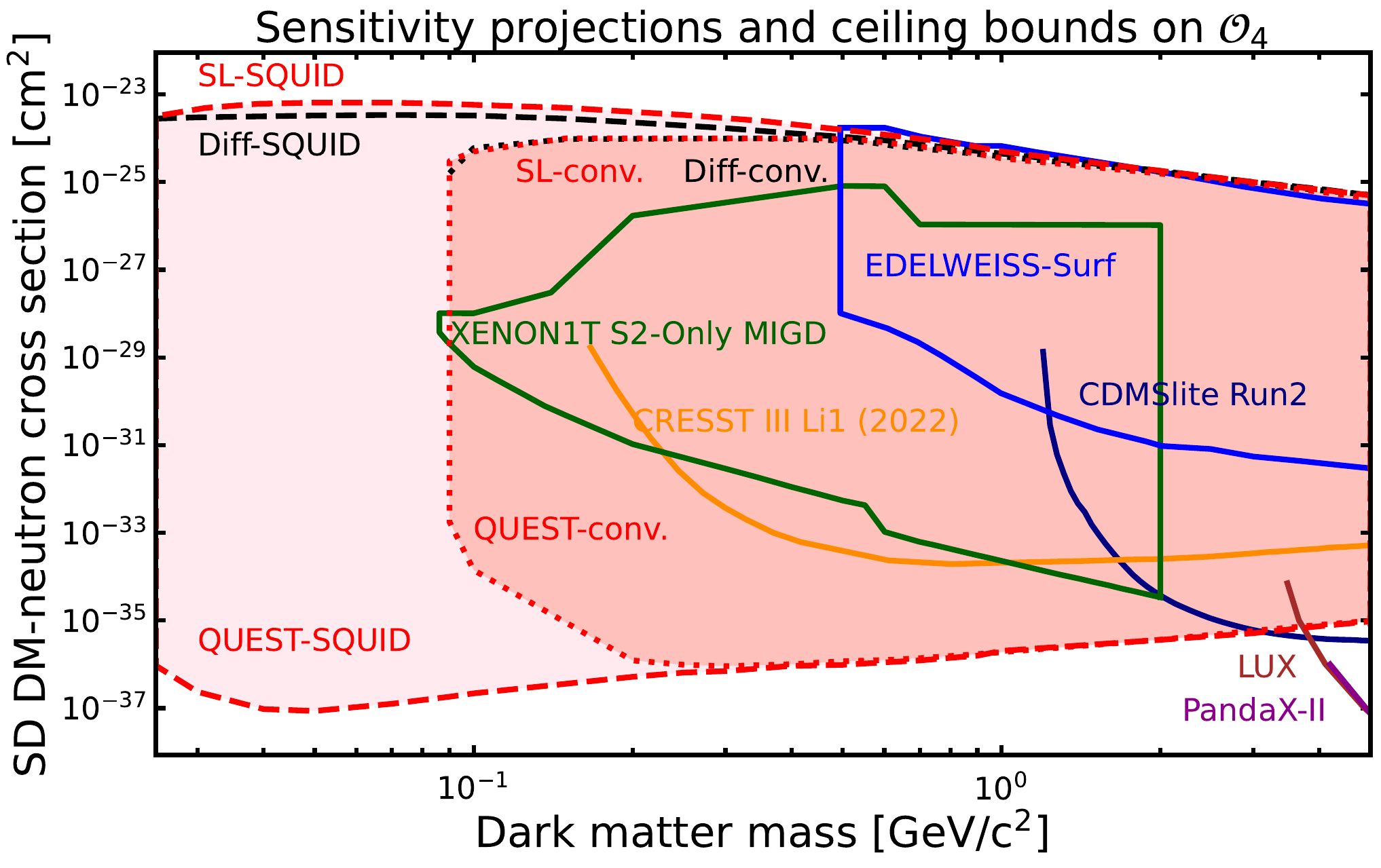}
\includegraphics[width=0.495\textwidth]{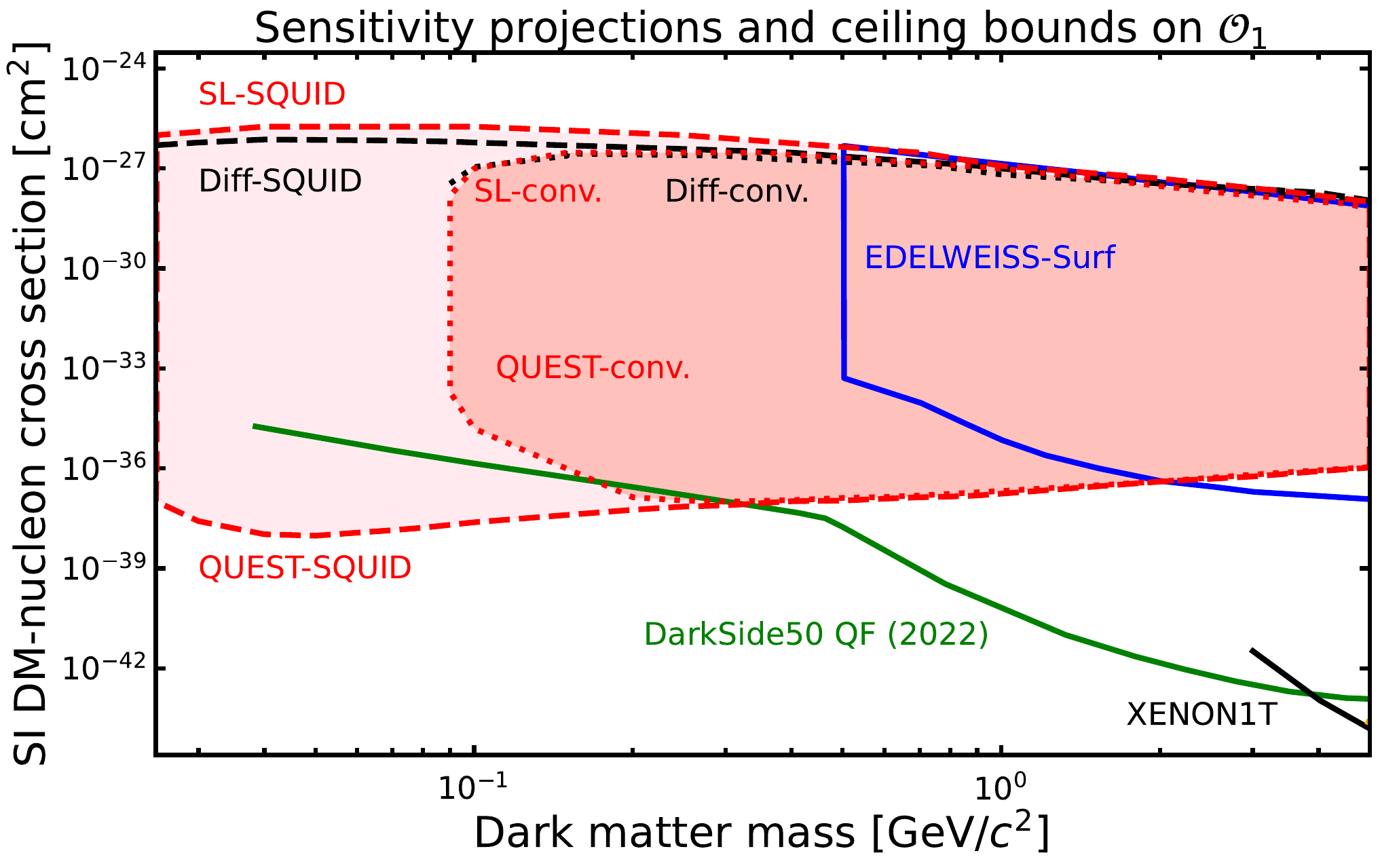}
\caption{The QUEST-DMC 90\%~C.L. limits on the cross-section for velocity and momentum independent operator $\mathcal{O}_1$ and $\mathcal{O}_4$ compared to the existing limits from Xenon 1T S2-only MIGD~\cite{XENON:2019zpr}, CRESST III (LiAlO$_2$)~\cite{CRESST_2022}, LUX (Xe)~\cite{LUXSD_2016}, CDMSlite (Ge)~\cite{CDMSLite_2018}, PandaX-II~\cite{PandaX-II:2018woa} and EDELWEISS~\cite{EDELWEISS:2019vjv} for SD and with existing limits from DarkSide-50~\cite{DarkSide-50:2022qzh}, XENON1T~\cite{XENON:2018voc}, and EDELWEISS~\cite{EDELWEISS:2019vjv} for SI. The upper limit sensitivity is given with the straight-line (SL) path and the diffusive (Diff) trajectories, depicted in red and black, respectively. The dashed lines correspond to SQUID-based readout systems, while the dotted lines denote conventional readout methods.}
\label{Ops_limits1}
\end{figure*}
Fig.~\ref{Ops_limits1} shows the projected 90\% C.L. sensitivity limits for canonical SI and SD interactions, represented by operators $\mathcal{O}_1$ and $\mathcal{O}_4$. 
The left panel illustrates sensitivity projections for the SD operator $\mathcal{O}_4$, compared to existing limits from Xenon 1T S2-only MIGD~\cite{XENON:2019zpr}, CRESST III (LiAlO$_2$)~\cite{CRESST_2022}, LUX~\cite{LUXSD_2016}, CDMSlite~\cite{CDMSLite_2018}, PandaX-II~\cite{PandaX-II:2018woa}, and EDELWEISS~\cite{EDELWEISS:2019vjv}.
For the SQUID readout, QUEST-DMC has sensitivity to exclude cross sections as low as $6.5 \times 10^{-24}\,{\rm cm}^2$ under the straight-line path model and $3.3 \times 10^{-24}\,{\rm cm}^2$ with the diffusion framework, within the $0.04$–$0.07$ GeV/$c^2$ mass range.
By contrast, with the conventional cold transformer readout, the sensitivity floor is $1 \times 10^{-24}\,{\rm cm}^2$, covering a higher mass range of $0.1$–$0.55$ GeV/$c^2$ for both propagation models. 
The right panel of Fig.~\ref{Ops_limits1} shows the projected 90\% C.L. sensitivity for SI interactions via operator $\mathcal{O}_1$, alongside current limits from DarkSide-50~\cite{DarkSide-50:2022qzh}, XENON1T~\cite{XENON:2018voc}, and EDELWEISS~\cite{EDELWEISS:2019vjv}. The mass range probed is similar to that for $\mathcal{O}_4$. 
\begin{figure*}[t]
\includegraphics[width=0.495\textwidth]{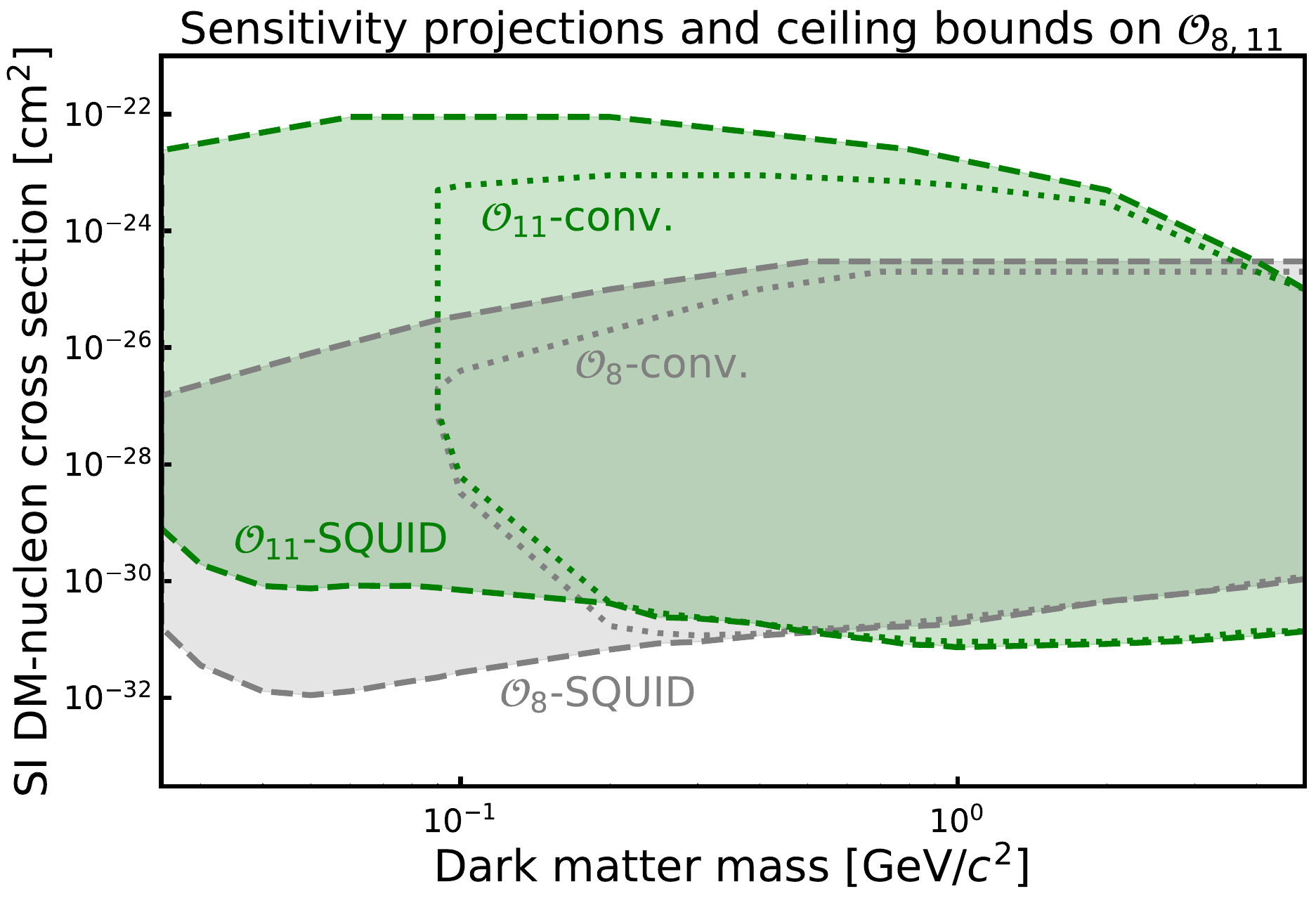}
\includegraphics[width=0.495\textwidth]{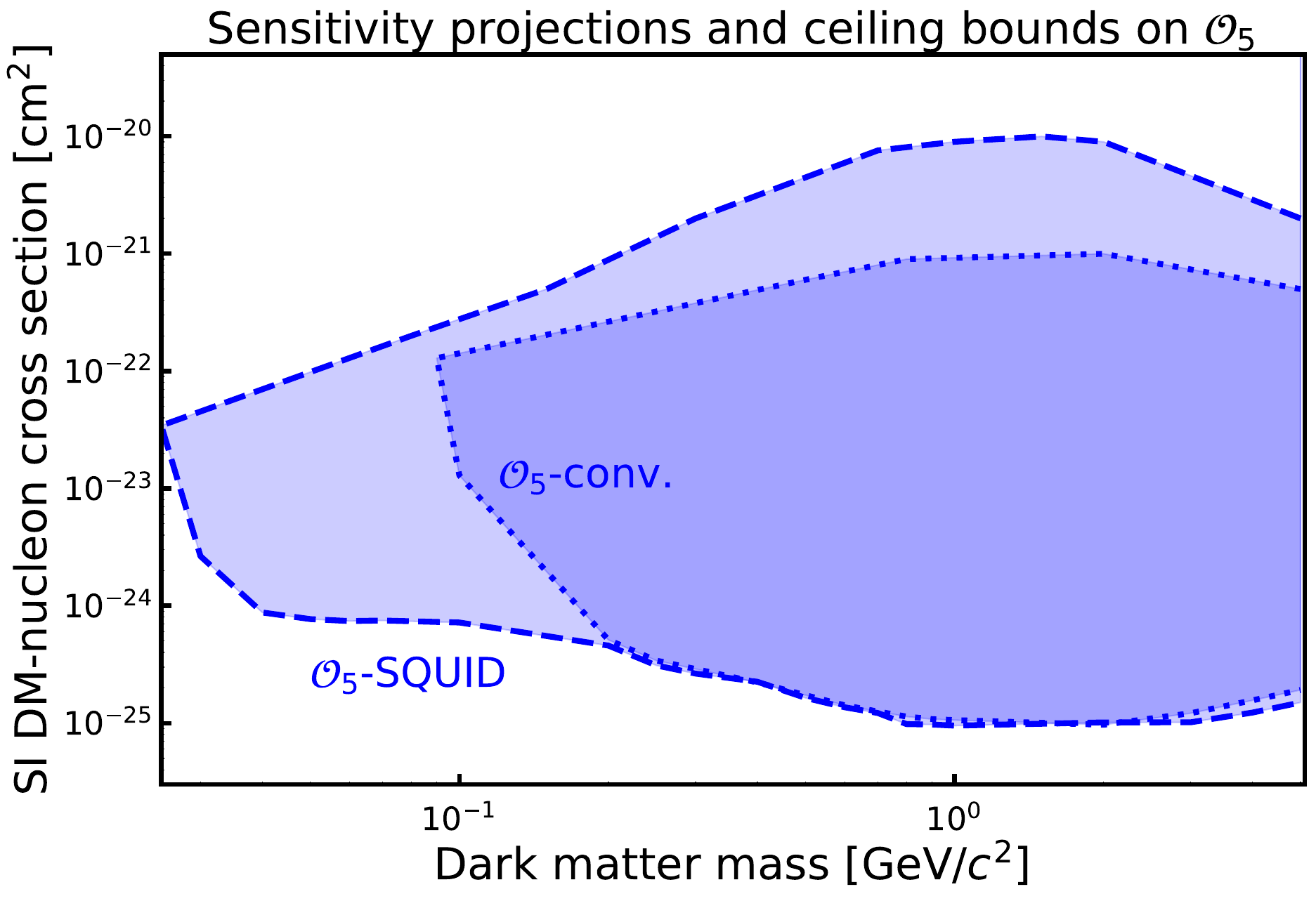}
\includegraphics[width=0.495\textwidth]{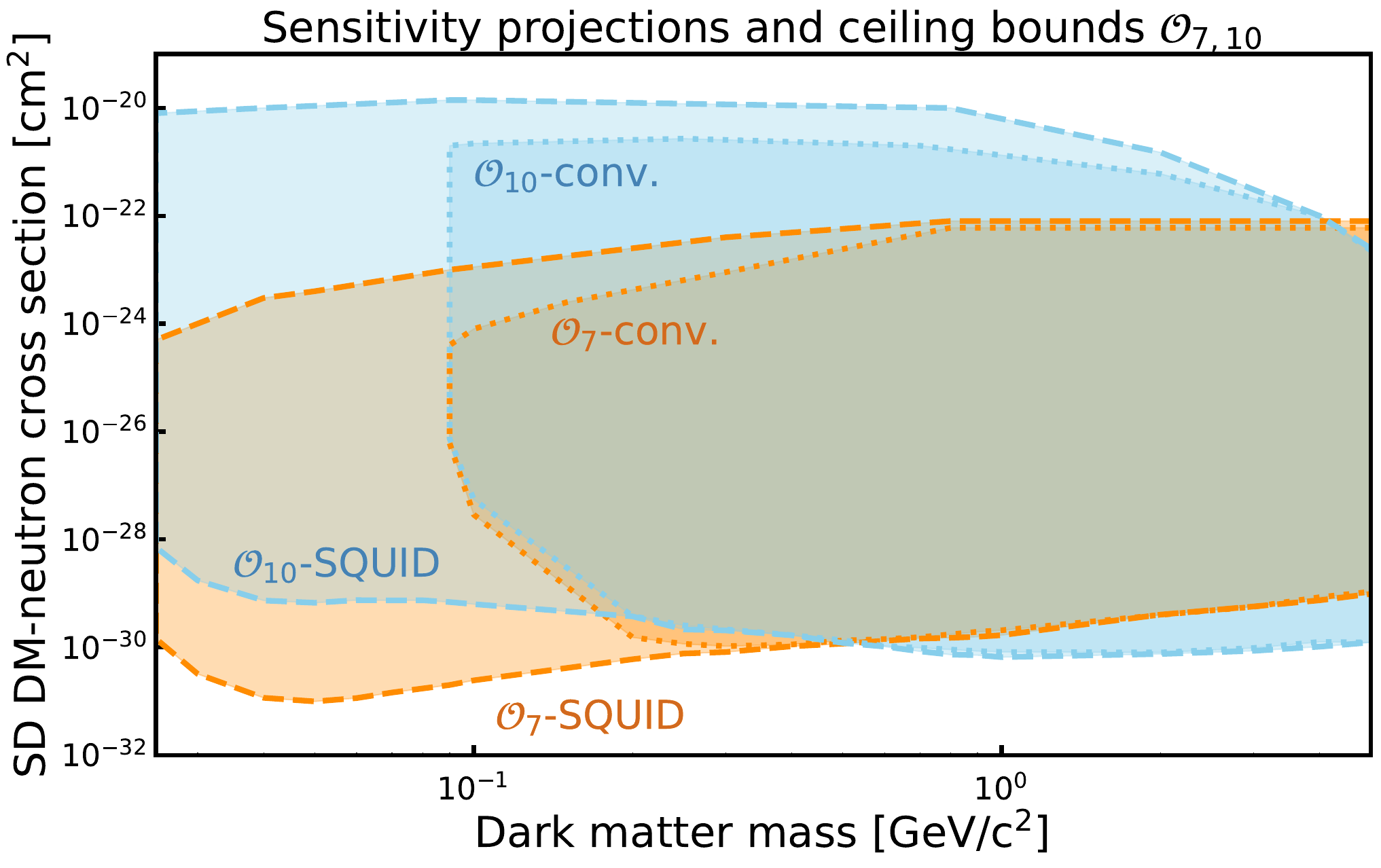}
\includegraphics[width=0.495\textwidth]{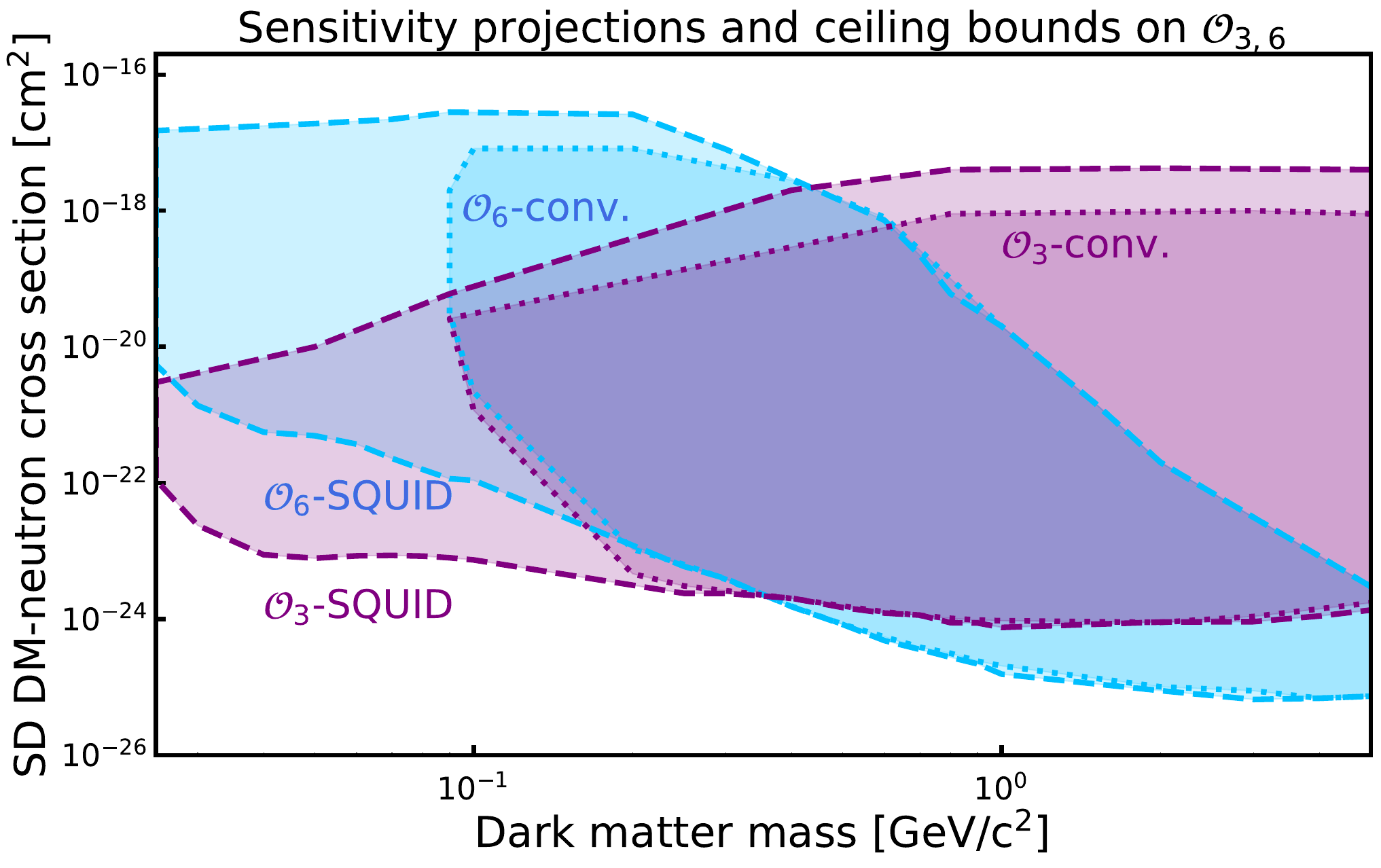}
\caption{ The QUEST-DMC 90\%~C.L. upper and lower exclusion limits on the cross-section sensitivity region for the non-relativistic EFT operators. The top panels show the SI operators $\mathcal{O}_{8,11}$ (left) and the strongest lower limit for the SI operator $\mathcal{O}_{5}$ (right). The bottom panels display the SD operators $\mathcal{O}_{7,10}$ (left) and $\mathcal{O}_{3,6}$ (right). Operators $\mathcal{O}_{9}$ and $\mathcal{O}_{12}$ are omitted from the plot due to their near-degeneracy with $\mathcal{O}_{10}$ and $\mathcal{O}_{7}$, respectively, in recoil spectra. These operators are numerically indistinguishable within the detector sensitivity range. Similarly, $\mathcal{O}_{13}$ and $\mathcal{O}_{14}$ closely match $\mathcal{O}_{3}$ up to scaling, owing to their shared dependence on both momentum and velocity. The dashed lines stand for the SQUID-based readout systems, and the dotted lines for the conventional readout.}
 \label{Ops_limits2}
\end{figure*}

Beyond these fundamental interactions, additional EFT operators probe different aspects of DM–nucleon interactions. In Fig.~\ref{Ops_limits2}, we present both the projected exclusion sensitivities and the corresponding sensitivity ceilings for the non-relativistic EFT operators. 
The top-left panel shows the SI operators $\mathcal{O}_{8}$ and $\mathcal{O}_{11}$, while the top-right panel displays the limits for the SI operator $\mathcal{O}_{5}$. The bottom-left panel presents the SD operators $\mathcal{O}_{7}$ and $\mathcal{O}_{10}$, and the bottom-right panel illustrates $\mathcal{O}_{3}$ and $\mathcal{O}_{6}$. 

The SI velocity-dependent operator $\mathcal{O}_8$ and the momentum-dependent operator $\mathcal{O}_{11}$ yield cross section constraints in the ranges $\sim [10^{-32} - 10^{-26}]\,{\rm cm}^2$ and $\sim [10^{-31} - 10^{-22}]\,{\rm cm}^2$, respectively. The results also constrain the largest cross sections for SI interactions, represented by the combined velocity- and momentum-dependent operator $\mathcal{O}_5$, which lies within $[10^{-25} - 10^{-20}]\,{\rm cm}^2$.

The projected exclusion sensitivity for the SD velocity-dependent operator $\mathcal{O}_7$ and the momentum-dependent operator $\mathcal{O}_{10}$ are within $[10^{-31} - 10^{-22}]\,{\rm cm}^2$ and $[10^{-29} - 10^{-20}]\,{\rm cm}^2$, respectively. Simultaneous velocity- and momentum-dependent interactions are captured by $\mathcal{O}_3$ and $\mathcal{O}_6$, whose lower-limit scale is larger than that of the SI operator $\mathcal{O}_5$. The constraints span several orders of magnitude, with the SQUID-based readout consistently achieving the strongest sensitivity. 

\begin{figure*}[t]
\includegraphics[width=0.495\textwidth]{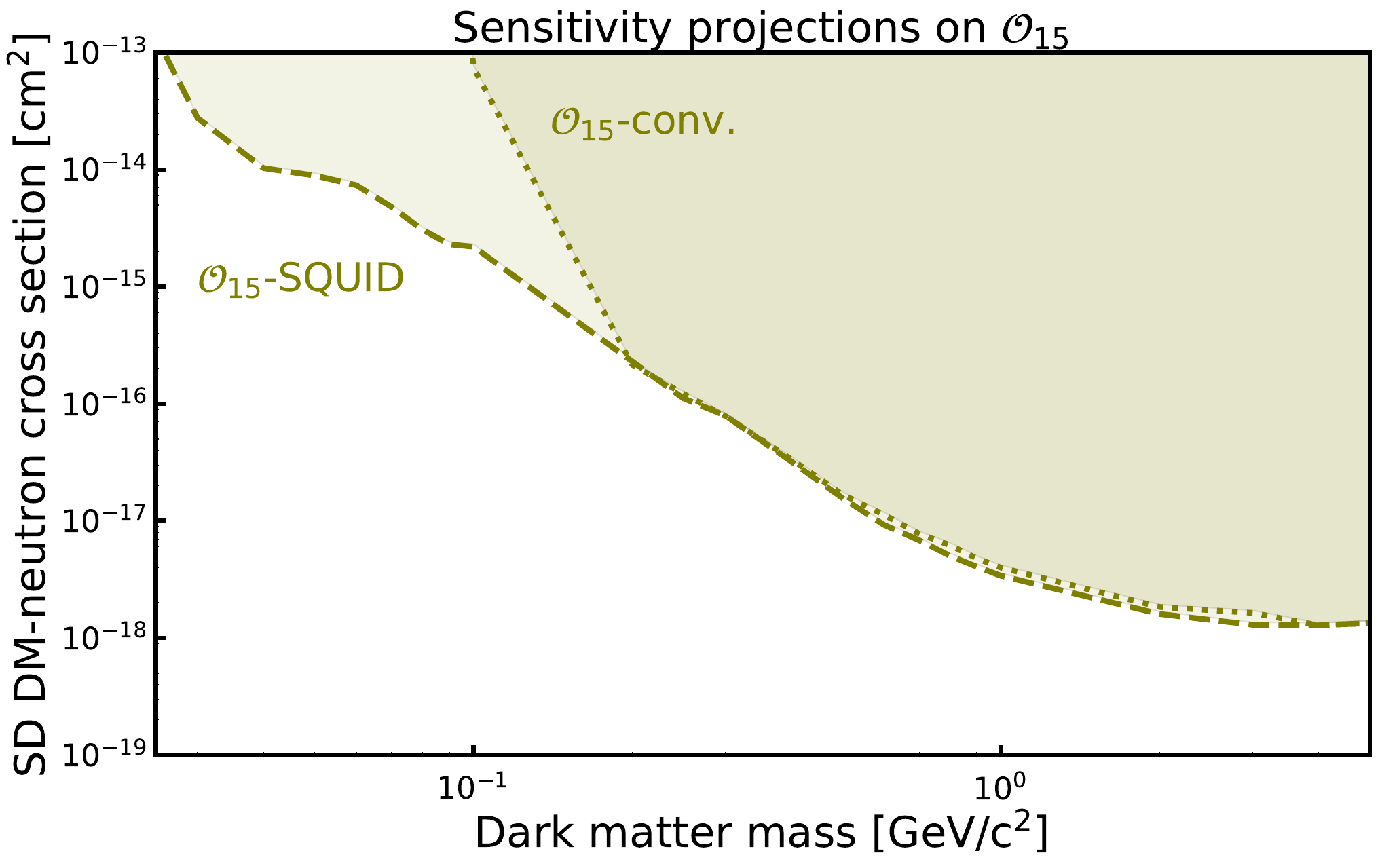}
\includegraphics[width=0.495\textwidth]{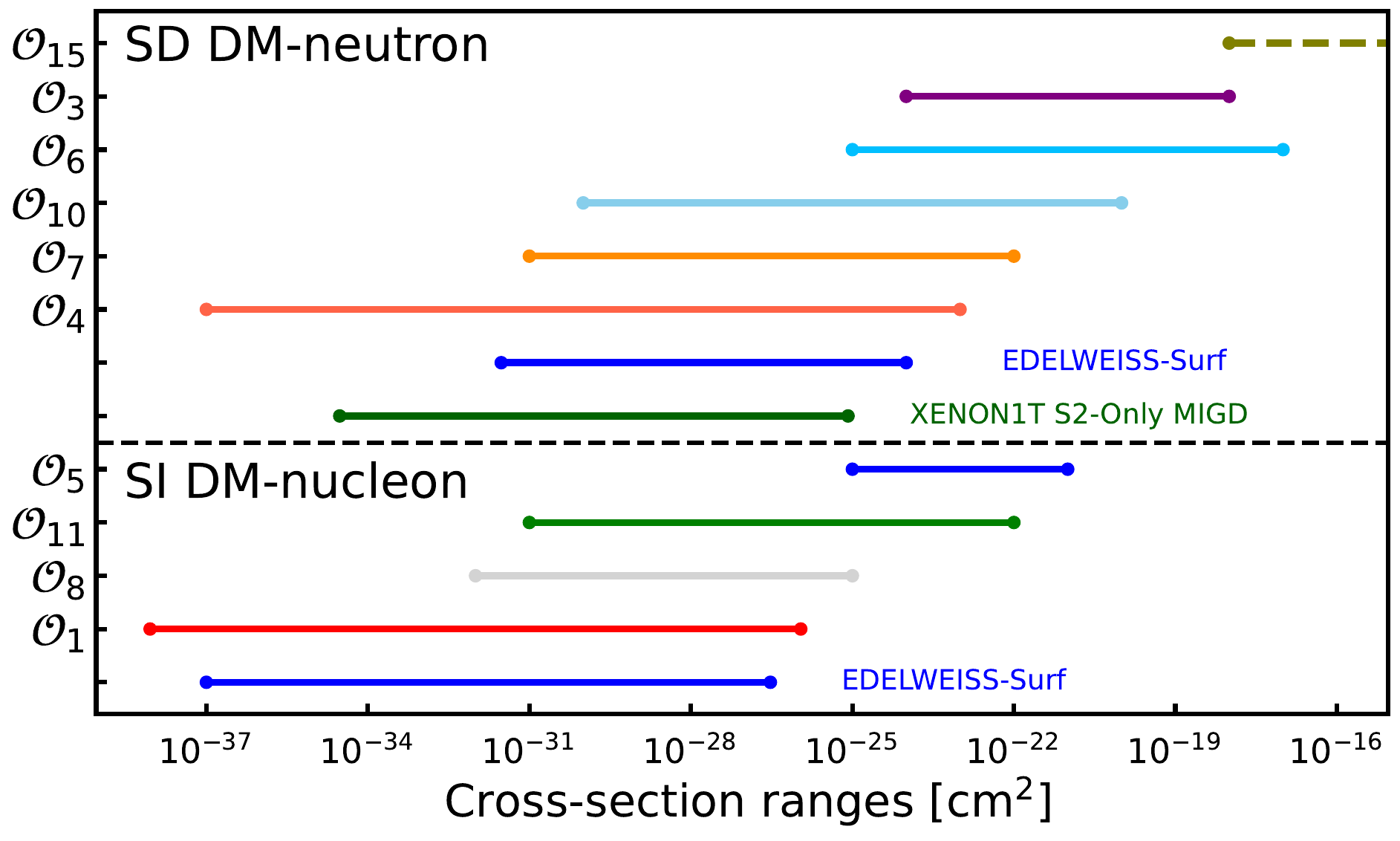}
\caption{The left panel shows QUEST-DMC 90\% C.L. lower bound on the highest cross-section for operator fifteen sets a threshold beyond which all parameter space is excluded, with no corresponding upper limit. The right panel provides an overview of the collective cross-section range for all analysed operators compared to the existing limits from Xenon 1T S2-only MIGD~\cite{XENON:2019zpr} (For SD) and EDELWEISS~\cite{EDELWEISS:2019vjv} (for SI and SD).}
 \label{Ops_limitsmax}
\end{figure*}

The event rate hierarchy across operators is shaped by their kinematic structure. Operators like $\mathcal{O}_1$ and $\mathcal{O}_4$, free from momentum or velocity suppression, yield the strongest signals. Intermediate cases, such as $\mathcal{O}_7$--$\mathcal{O}_{12}$, include linear dependencies on either variable. Stronger suppression occurs for operators like $\mathcal{O}_3$ and $\mathcal{O}_6$, involving products of velocity and momentum or higher powers of momentum transfer. $\mathcal{O}_5$, $\mathcal{O}_{13}$, and $\mathcal{O}_{14}$ also exhibit significant signal suppression due to their mixed dependence. $\mathcal{O}_{15}$ remains the least sensitive operator due to its extreme suppression.
The left panel of Fig.~\ref{Ops_limitsmax} illustrates this extremal case of an SD operator with suppression arising from both the squared momentum transfer and the transverse velocity. Its recoil signature is significantly weaker than that of other operators, leading to exclusion limits well above the rest of the EFT space. Atmospheric attenuation further reduces the high-velocity DM population, making such a signal nearly undetectable even under favourable conditions.

Degeneracies arise between operators that produce similar recoil spectra, resulting in nearly indistinguishable signals within the detector energy sensitivity range. In particular, operators $\mathcal{O}_{13}$ and $\mathcal{O}_{14}$ yield almost identical responses, while pairs such as $\mathcal{O}_9$ and $\mathcal{O}_{10}$, and $\mathcal{O}_7$ and $\mathcal{O}_{12}$, differ only by a small normalisation factor.
 
The right panel of Fig.~\ref{Ops_limitsmax} summarises the full range of cross sections probed for each EFT operator using the SQUID-based QUEST-DMC configuration, together with results from the literature. Horizontal bars indicate the span from the lowest projected sensitivity to the highest excluded value, while a dashed line denotes the unconstrained upper limit for $\mathcal{O}_{15}$. This shows the breadth of EFT parameter space covered by QUEST-DMC, particularly in the low-mass and suppressed-interaction regimes.

\subsection{Mapping Projected Sensitivities to Relativistic EFT Couplings} \label{subsec:rel-couplings}
To connect experimental constraints to UV-complete models, we express the non-relativistic EFT operators $\mathcal{O}_i$ as low-energy limits of relativistic bilinear interactions between DM and nucleons. 
This approach enables a broader exploration of DM–nucleon and DM–neutron interactions beyond the standard SI and SD interpretation. Tables~\ref{tab:DM_SI}–\ref{tab:DM_tensor} present the mapping between non-relativistic EFT operators and the corresponding relativistic structures, expressed in terms of Dirac bilinears $\bar{\chi} \Gamma \chi$ for DM and $\bar{N} \Gamma' N$ for nucleons, where $\Gamma, \Gamma' \in \{1,\, \gamma^5,\, \gamma^\mu \gamma^5,\, i \sigma^{\mu\nu} q_\nu\}$ represent scalar, pseudoscalar, axialvector, and tensor interactions, respectively. In these tables, the bilinear combinations yield linear combinations or rescaled forms of the Wilson coefficients of the non-relativistic operators; for example, an axialvector-axialvector interaction maps to $|4 \,c_4|$ in the case of operator $\mathcal{O}_4$.

In classifying the relativistic interactions presented in Tables~\ref{tab:DM_SI}–\ref{tab:DM_tensor}, we group operators according to the spinor structure of the nucleon and DM bilinears (e.g., scalar, pseudoscalar, axialvector, or tensor), regardless of the full Lorentz structure of the total bilinear product. For example, terms involving momentum insertions such as $K^\mu(\bar{N}N)$, which transform as a vector overall, are categorised under scalar nucleon currents due to the presence of the spinor structure $\bar{N}N$. 

Table~\ref{tab:DM_SI} lists operators generated by scalar currents on the nucleon side, combined with various DM bilinears, including scalar, pseudoscalar, axialvector, and tensor structures. These combinations primarily contribute to SI interactions. All momentum-dependent terms are normalised by a reference mass scale, taken here to be the nucleon mass $M = 0.938$ GeV, ensuring dimensional consistency across the operator expressions.

Table~\ref{tab:DM_SD} presents operators arising from axialvector currents on the nucleon side, paired with various DM bilinears. Table~\ref{tab:DM_SD2} focuses on pseudoscalar nucleon currents, which similarly lead to SD-type structures. Table~\ref{tab:DM_tensor} summarises interactions involving tensor currents on the nucleon side; combinations with scalar or pseudoscalar DM currents yield mixed operator types, while axialvector and tensor DM currents contribute additional SD terms. All expressions are given in a normalised form consistent with the conventions of~\cite{Anand:2013yka}. 

\begin{table}[t]
\centering
\renewcommand{\arraystretch}{2}
\begin{tabular}{p{5.5cm} c}
\hline
Relativistic Operator \newline 
\small (Spinor Bilinear Labelling)
& Factor $\times \mathcal{O}_i$ \\
\hline\hline
$(\bar{\chi} \chi)(\bar{N} N)$ \newline \small
Scalar–Scalar & $\mathcal{O}_1$ [and $4 \dfrac{m_\chi m_N}{M^2} \mathcal{O}_1$] \\ \hline
$(\bar{\chi} i \sigma^{\mu\nu} \dfrac{q_\nu}{M} \chi) \dfrac{K_\mu}{M} (\bar{N} N)$ \newline \small
Tensor–Scalar & $\displaystyle{1 \over M^2}(\dfrac{m_N}{ m_\chi} \vec{q}^2 \mathcal{O}_1 - 4 m_N^2 \mathcal{O}_5)$ \\ \hline
$(\bar{\chi} \gamma^\mu \gamma^5 \chi) \dfrac{K_\mu}{M} (\bar{N} N)$ \newline \small
Axialvector–Scalar & $4 \dfrac{m_N}{M} \mathcal{O}_8$ \\ \hline
$i(\bar{\chi} \gamma^5 \chi)(\bar{N} N)$ \newline \small
Pseudoscalar–Scalar & $-\dfrac{m_N}{m_\chi} \mathcal{O}_{11}$ [and $-4 \dfrac{m_N^2}{M^2} \mathcal{O}_{11}$] \\
\hline
\end{tabular}
\caption{SI DM–nucleon effective interactions with scalar, tensor, axialvector, and pseudoscalar currents on the DM side ($\bar{\chi} \Gamma \chi$, where $\Gamma \in \{1,\, \gamma^5,\, \gamma^\mu \gamma^5,\, i \sigma^{\mu\nu} q_\nu\}$ denotes the corresponding Lorentz bilinear), and scalar current on the nucleon side ($\bar{N} N$). Operators are named according to the spinor bilinears before momentum contractions. 
The terms enclosed in brackets in the second column represent momentum-suppressed contributions arising from four-momentum insertions on both the DM and nucleon sides. The normalisation and conversion to the non-relativistic basis match the results of Ref.~\cite{Anand:2013yka}. }
\label{tab:DM_SI}
\end{table}
\vspace{0.8em}
\begin{table}[t]
\centering
\renewcommand{\arraystretch}{1.9}
\begin{tabular}{p{5.5cm} c}
\hline
Relativistic Operator \newline
\small (Spinor Bilinear Labelling)
& ~~~~~~~~~~~Factor $\times \mathcal{O}_i~~~~~~~~~$ \\
\hline\hline
$(\bar{\chi} \gamma^\mu \gamma^5 \chi)(\bar{N} \gamma_\mu \gamma^5 N)$ \newline \small Axialvector–Axialvector &~~~~~ $-4\, \mathcal{O}_4$ \\ \hline
$\dfrac{P^\mu}{M}(\bar{\chi} \chi) (\bar{N} \gamma_\mu \gamma^5 N)$ \newline \small Scalar–Axialvector &~~~~~ $-4\, \dfrac{m_\chi}{M} \mathcal{O}_7$ \\ \hline
$(\bar{\chi} i \sigma^{\mu\nu} \dfrac{q_\nu}{M} \chi)(\bar{N} \gamma_\mu \gamma^5 N)$ \newline \small Tensor–Axialvector &~~~~~ $-4\, \dfrac{m_N}{M} \mathcal{O}_9$ \\ \hline
$i \dfrac{P^\mu}{M}(\bar{\chi} \gamma^5 \chi) (\bar{N} \gamma_\mu \gamma^5 N)$ \newline \small Pseudoscalar–Axialvector &~~~~~ $4\, \dfrac{m_N}{M} \mathcal{O}_{14}$ \\
\hline
\end{tabular}
\caption{Same as Table~\ref{tab:DM_SI} but for SD interaction and axialvector current on the nucleon side ($\bar{N} \gamma^\mu \gamma^5 N$). Classification is based on the spinor bilinears only; momentum factors are treated separately. Our normalisations match those of Ref.~\cite{Anand:2013yka}.}
\label{tab:DM_SD}
\end{table}
\vspace{0.8em}
\begin{table}[t]
\centering
\renewcommand{\arraystretch}{1.9}
\begin{tabular}{p{5.5cm} c}
\hline
Relativistic Operator \newline
\small (Spinor Bilinear Labelling)
& Factor $\times \mathcal{O}_i$ \\
\hline\hline
$i(\bar{\chi} \chi)(\bar{N} \gamma^5 N)$ \newline \small
Scalar–Pseudoscalar & $\mathcal{O}_{10}$ [and $4 \dfrac{m_\chi m_N}{M^2} \mathcal{O}_{10}$] \\
\hline
$(\bar{\chi} \gamma^5 \chi)(\bar{N} \gamma^5 N)$ \newline \small
Pseudoscalar–Pseudoscalar & $-\dfrac{m_N}{m_\chi} \mathcal{O}_6$ [and $-4 \dfrac{m_N^2}{M^2} \mathcal{O}_6$] \\
\hline
$i(\bar{\chi} \gamma^\mu \gamma^5 \chi) \dfrac{K_\mu}{M} (\bar{N} \gamma^5 N)$ \newline \small
Axialvector–Pseudoscalar & $4 \dfrac{m_N}{M} \mathcal{O}_{13}$ \\
\hline
$i(\bar{\chi} i \sigma^{\mu\nu} \dfrac{q_\nu}{M} \chi) \dfrac{K_\mu}{M} (\bar{N} \gamma^5 N)$ \newline \small
Tensor–Pseudoscalar & \hspace{-0.4cm}$\displaystyle{4 \vec{q}^{\,2} \over M^2}(\displaystyle{m_N \over 4 m_\chi} \mathcal{O}_{10}+\mathcal{O}_{12})$+$\displaystyle{4 m_N^2 \over M^2} \mathcal{O}_{15}$ \\
\hline
\end{tabular}
\caption{Same as Table~\ref{tab:DM_SD}, but with the pseudoscalar current on the nucleon side ($\bar{N}\gamma^5 N$) and normalisations match those reported in Ref.~\cite{Anand:2013yka}.}
\label{tab:DM_SD2}
\end{table}
\vspace{0.8em}
\begin{table}[t]
\centering
\renewcommand{\arraystretch}{2}
\begin{tabular}{p{5.5cm} c}
\hline
Relativistic Operator \newline
\small (Spinor Bilinear Labelling)
& Factor $\times \mathcal{O}_i$ \\
\hline\hline
$\dfrac{P^\mu}{M}(\bar{\chi} \chi) (\bar{N} i\sigma_{\mu\alpha} \dfrac{q^\alpha}{M} N)$ \newline \small
Scalar–Tensor & $\displaystyle{ m_\chi m_N\over M^2}(-\dfrac{\vec{q}^{\,2}}{ m_N^2} \mathcal{O}_1 + 4 \mathcal{O}_3$) \\
\hline
$i\dfrac{P^\mu}{M} (\bar{\chi} \gamma^5 \chi) (\bar{N} i\sigma_{\mu\alpha} \dfrac{q^\alpha}{M} N)$ \newline \small
Pseudoscalar–Tensor & $\displaystyle{1 \over M^2}(\vec{q}^{\,2} \mathcal{O}_{11} + 4 m_N^2 \mathcal{O}_{15})$\\
\hline
$(\bar{\chi} \gamma^\mu \gamma^5 \chi)(\bar{N} i\sigma_{\mu\alpha} \dfrac{q^\alpha}{M} N)$ \newline \small
Axialvector–Tensor & $4 \dfrac{m_N}{M} \mathcal{O}_9$ \\
\hline
$(\bar{\chi} i \sigma^{\mu\nu} \dfrac{q_\nu}{M} \chi)(\bar{N} i\sigma_{\mu\alpha} \dfrac{q^\alpha}{M} N)$ \newline \small
Tensor–Tensor & $\displaystyle{4 \over M^2}( \vec{q}^{\,2}\mathcal{O}_{4} -m_N^2 \mathcal{O}_{6})$ \\
\hline
\end{tabular}
\caption{Tensor bilinear on the nucleon side ($\bar{N} \sigma_{\mu\alpha} q^\alpha N$) combined with scalar and pseudoscalar currents from the DM side leads to mixed SI and SD operators, while axialvector and tensor on the DM side correspond to SD operators (see Ref.~\cite{Anand:2013yka}).}
\label{tab:DM_tensor}
\end{table} 

\begin{figure*}[t]
\centering
\includegraphics[width=0.495\textwidth]{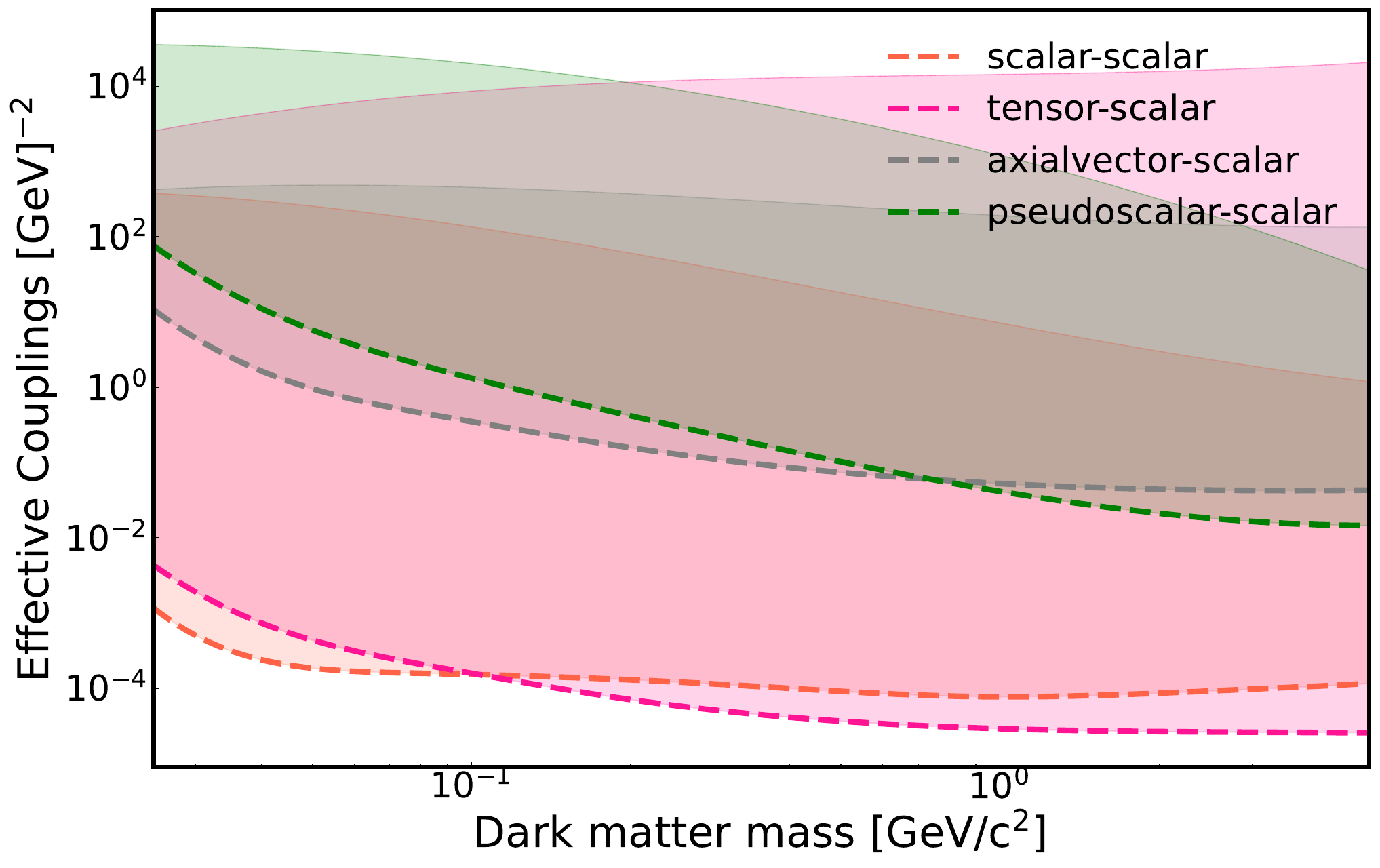}
\includegraphics[width=0.495\textwidth]{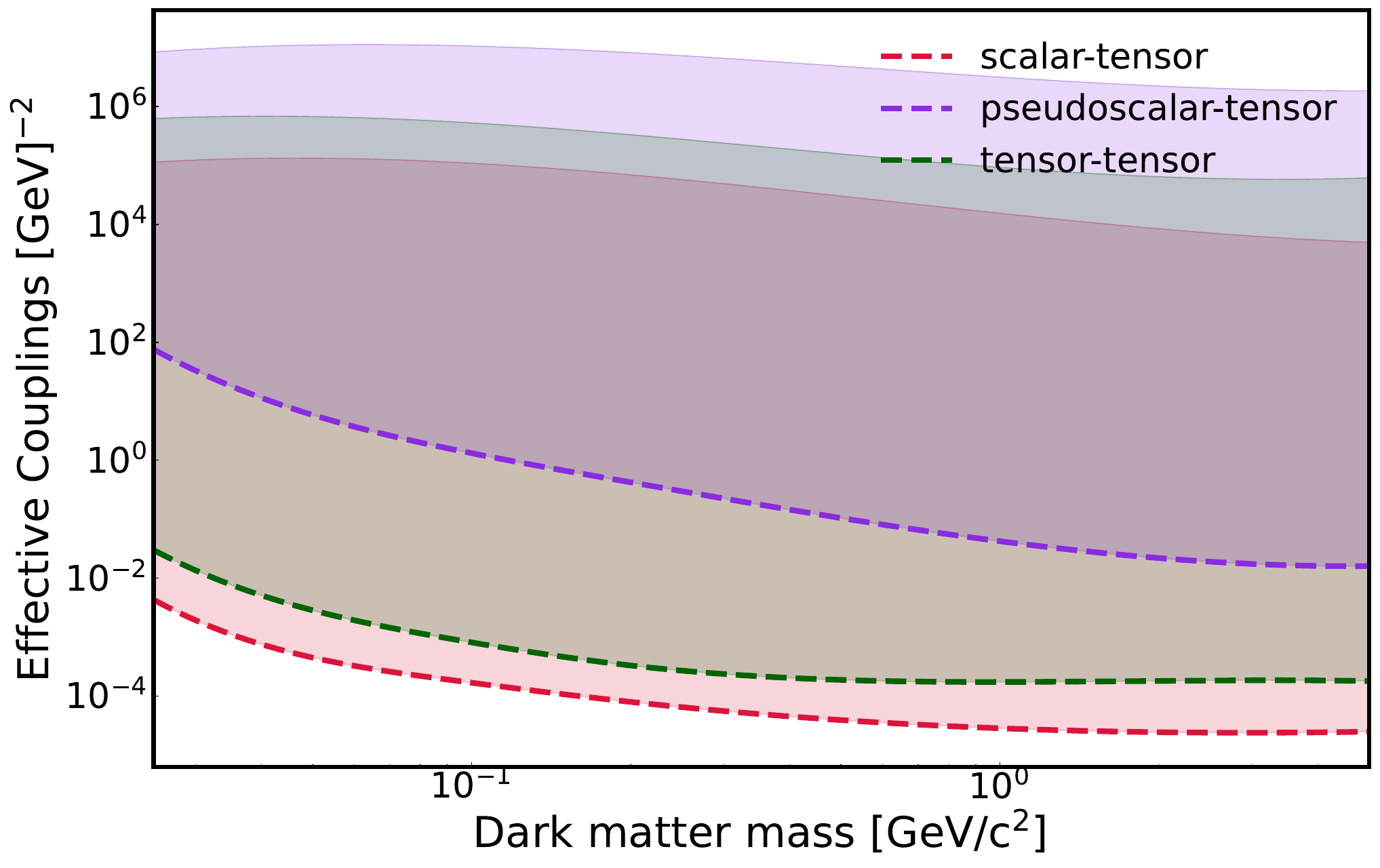}
\includegraphics[width=0.495\textwidth]{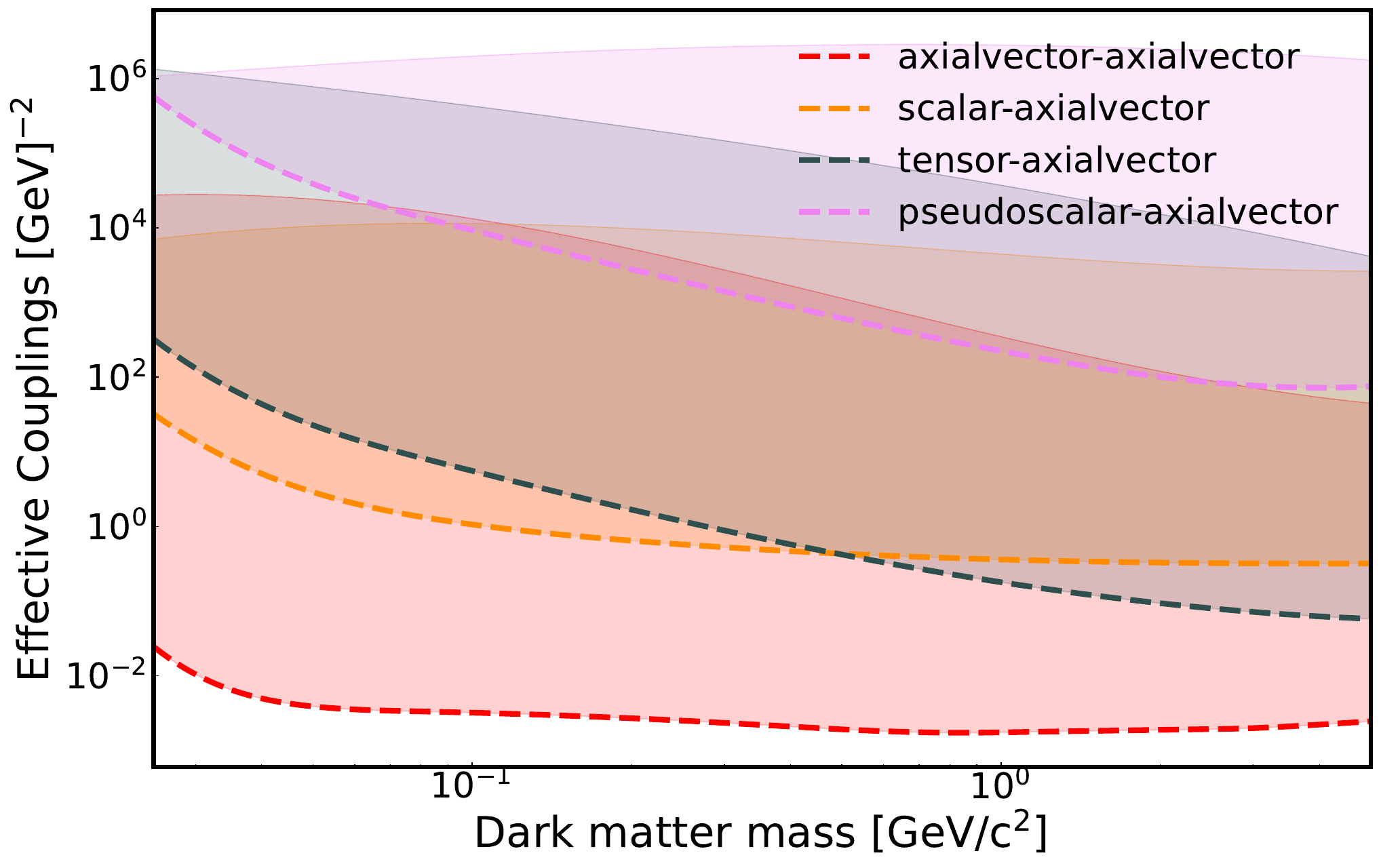}
\includegraphics[width=0.495\textwidth]{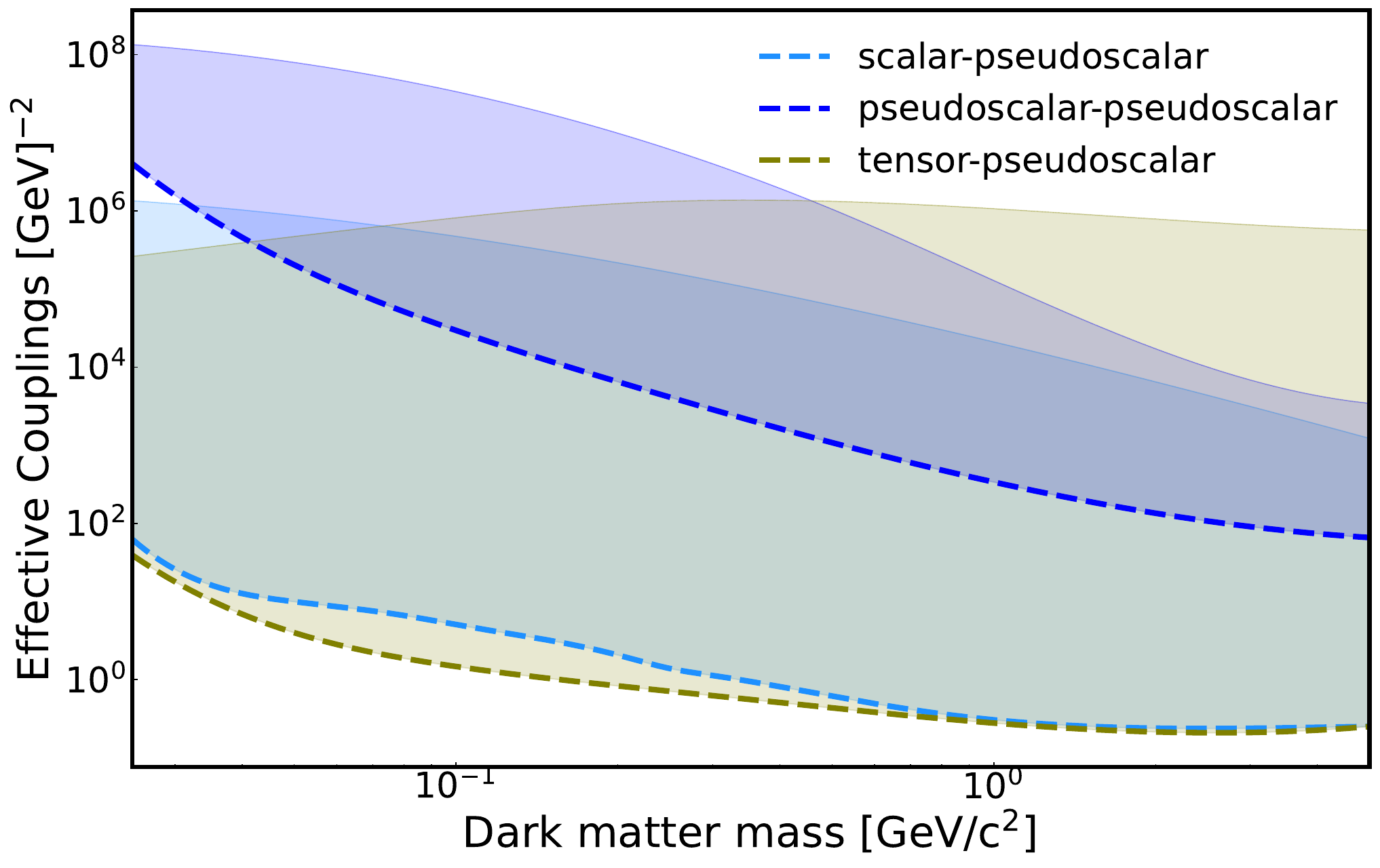}
\caption{
Projected sensitivity to effective couplings for a range of DM–nucleon interactions classified by the Lorentz bilinears in the relativistic theory, shown for the SQUID-based readout. Each panel corresponds to a different class of nucleon bilinears: scalar (top left), tensor (top right), axialvector (bottom left), and pseudoscalar (bottom right). The colored shaded bands span the region between the lowest point of the projected coupling sensitivity and the highest point of the sensitivity ceiling. The couplings are matched to non-relativistic EFT operators as detailed in Tables~\ref{tab:DM_SI}–\ref{tab:DM_tensor}.}
\label{fig:EFT_couplings}
\end{figure*}

\begin{figure}[t]
\centering
\includegraphics[width=0.6\textwidth]{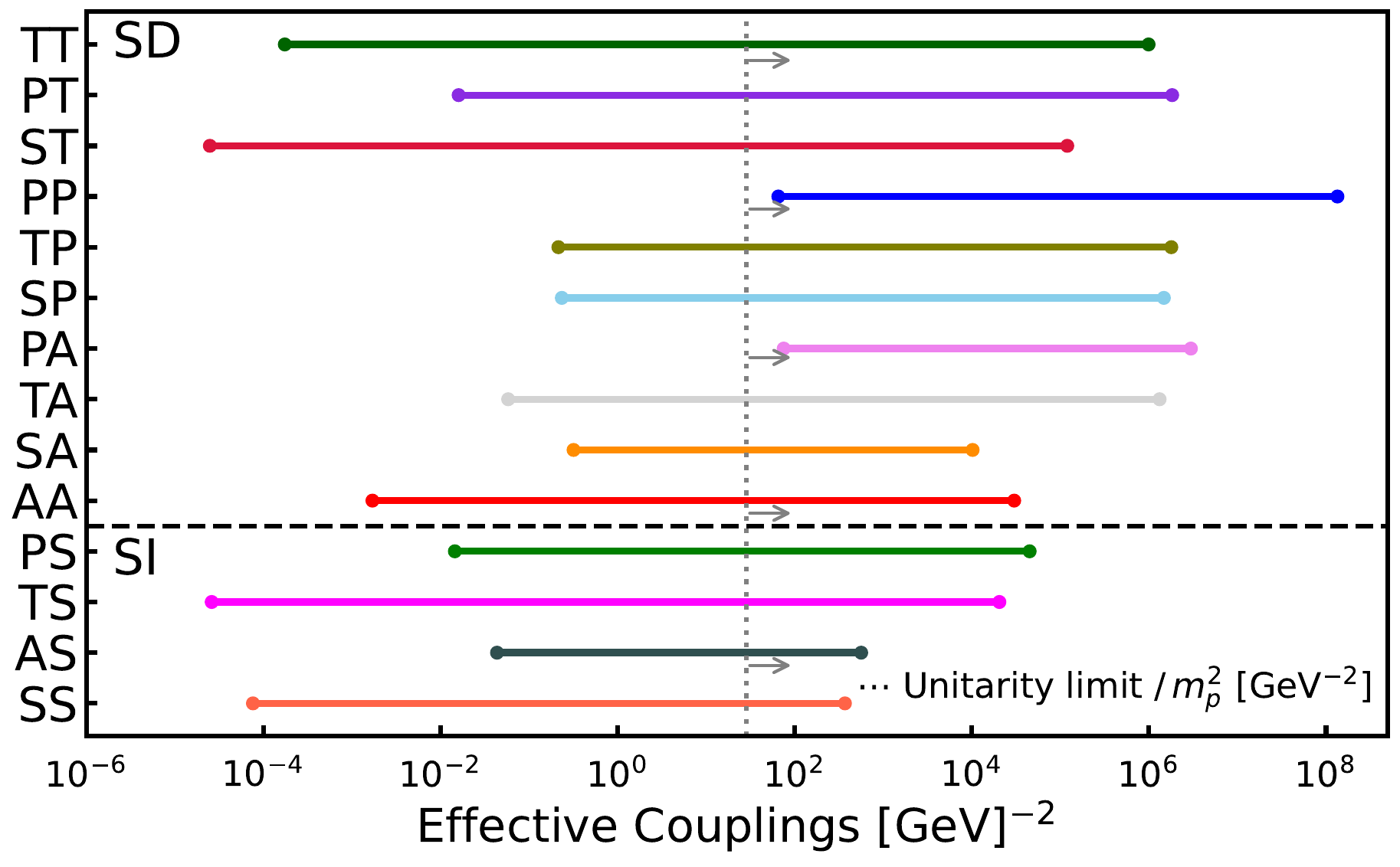}
\caption{
Effective coupling ranges for DM–nucleon and DM–neutron interactions, grouped by their underlying relativistic bilinear types (SS, TS, AA, etc.) as defined in Tables~\ref{tab:DM_SI}–\ref{tab:DM_tensor}. The couplings represent rescaled Wilson coefficients derived from the QUEST-DMC SQUID-based sensitivity projections shown in Fig.~\ref{fig:EFT_couplings}. The vertical dashed line with right-pointing arrows indicates the region above the unitarity limit. 
}
\label{fig:summary_limits}
\end{figure}

To obtain projected exclusion sensitivities on the full set of relativistic DM–nucleon couplings, we construct the complete non-relativistic operator combinations corresponding to each relativistic bilinear structure, following the standard mappings from Lorentz-invariant currents to Galilean-invariant EFT operators, as summarised in Tables~\ref{tab:DM_SI}–\ref{tab:DM_tensor}. These combinations typically involve multiple operators with distinct momentum and velocity dependencies and are incorporated directly into the differential recoil rate calculation. Two sets of rates are computed, one assuming unattenuated DM flux to determine the experiment's projected sensitivity floor, and another including atmospheric attenuation, which limits the flux of high-cross-section DM and defines the sensitivity ceiling. The resulting spectra are then propagated through the full profile likelihood ratio analysis pipeline, which accounts for detector response modelling, energy threshold effects, and expected backgrounds. This procedure preserves the full kinematic structure of each interaction and ensures that spectral features, such as recoil suppression from velocity- or momentum-dependent operators, and distortions arising from broad kinematic distributions, are accurately captured. The inclusion of background expectations enables realistic limit setting under finite exposure, particularly for suppressed interactions near the threshold. Consequently, the projected limits on EFT couplings presented here reflect the detector-level phenomenology of each interaction type, enabling direct comparison across operator classes and consistent interpretation within UV-complete models.

Fig.~\ref{fig:EFT_couplings} illustrates the projected sensitivity floors and ceilings on effective couplings for sub-GeV DM interactions, covering the full set of interaction structures defined in Tables~\ref{tab:DM_SI}–\ref{tab:DM_tensor}. We note that for certain interactions, the corresponding non-relativistic operator scaling is indicated in brackets in these tables, reflecting the same spinor structure but with additional momentum-suppressed contributions from four-momentum insertions on both the DM and nucleon sides; however these contributions are omitted from the plots as they do not represent independent physical operators.
The PandaX-II experiment~\cite{PandaX-II:2018woa} has placed lower bounds on effective couplings for WIMP masses above $\sim 5~{\rm GeV}/c^2$ which are not shown in Fig.~\ref{fig:EFT_couplings} as our analysis focuses on the sub-GeV mass regime.

Notably, the projected sensitivity floors presented here indicate that Lorentz structures involving tensor–scalar and scalar–tensor spinor combinations can lead to more stringent sensitivities than the conventionally dominant scalar–scalar and axialvector–axialvector forms. The tensor–scalar case arises from the combination of $\mathcal{O}_1$ and $\mathcal{O}_5$ (both SI), while the scalar–tensor structure maps onto a mix of $\mathcal{O}_1$ and $\mathcal{O}_3$, incorporating both SI and SD contributions. Across the sub-GeV mass range, tensor–scalar interactions yield stronger projected coupling sensitivities than scalar–scalar for DM masses above 0.1\,GeV; for example, at $m_\chi = 1\,{\rm GeV}/c^2$, the tensor–scalar limit is $3.0 \times 10^{-5}\,{\rm GeV}^{-2}$, compared to $7.7 \times 10^{-5}\,{\rm GeV}^{-2}$ for scalar–scalar. Scalar–tensor spinor combinations consistently yield tighter limits than axialvector–axialvector couplings, reaching $2.84 \times 10^{-5}\,{\rm GeV}^{-2}$ at $1~{\rm GeV}/c^2$ versus $1.83 \times 10^{-3}\,{\rm GeV}^{-2}$. This difference becomes more significant at higher masses. Among all interaction types, the pseudoscalar–pseudoscalar structure, corresponding to the non-relativistic operator $\mathcal{O}_6$, exhibits the highest sensitivity ceiling, consistent with the sharp cross-section growth in the low-mass region as observed in the lower-right panel of Fig.~\ref{Ops_limits2}.

A broader overview is provided in Fig.~\ref{fig:summary_limits}, which summarises the constrained effective coupling ranges across all interaction classes. These are grouped by relativistic bilinear combinations, such as scalar-scalar (SS), tensor-scalar (TS), axialvector-axialvector (AA) etc. The horizontal bars span the coupling range from the projected exclusion sensitivity floor to the sensitivity ceiling imposed by atmospheric attenuation. Only results from the SQUID-based readout are shown, as they represent the most stringent projected constraints.

The observed variation in coupling sensitivity across operator classes, in some cases as much as two orders of magnitude, highlights the importance of a systematic and comprehensive mapping between relativistic bilinears and non-relativistic EFT operators. Rather than relying solely on conventional benchmarks such as SI or SD interactions, this framework accounts for all Lorentz structures that can map onto non-relativistic operators with recoil spectra within the energy sensitivity range of a given detector. These mappings help identify which interactions are more likely to evade or dominate experimental sensitivity, depending on their momentum and velocity dependence, thereby enabling a more accurate connection between low-energy phenomenology and UV-complete theoretical models. However, in realistic UV-complete theories, it is uncommon for a single relativistic DM–nucleon operator to arise in isolation. Interactions are typically generated at the DM–quark/gluon level and, after hadronisation and matching onto nucleon-level degrees of freedom, give rise to combinations of both relativistic and non-relativistic operators. Our analysis in this section treats individual relativistic bilinears as a phenomenologically useful intermediate step between the non-relativistic EFT framework and UV-complete models. A full treatment, beginning with DM–quark/gluon EFT operators and their systematic matching onto nucleons (see e.g. Ref.~\cite{Bishara:2017pfq}), would more faithfully reflect the operator structure predicted by UV completions. While such an approach lies beyond the scope of the present study, it represents a promising direction for future work.

\section{Summary and Conclusions}

The interactions between DM and Standard Model particles are commonly formulated within the EFT framework. EFT provides a systematic, model-independent approach to parametrising all operators consistent with the relevant symmetries at a given energy scale, typically organised by operator dimension. These operators may encode SI and SD interactions, as well as velocity-dependent or momentum-suppressed terms, offering a broader and more general perspective than traditional SI and SD analyses alone. By expressing DM–nucleon interactions in terms of EFT operators, one can map UV-complete theories onto experimentally accessible observables, such as recoil energy spectra and interaction cross sections. Constraints derived from experimental data can thus be translated into upper bounds on operator coefficients, providing a robust and flexible means of testing a wide variety of DM models.

Here we project the sensitivity of the QUEST-DMC experiment to DM interactions within the framework of non-relativistic EFT, particularly in the sub-GeV mass range. Leveraging the unique properties of superfluid He-3 and the precision of quantum sensing technologies, the experiment is projected to achieve exceptional sensitivity to both SI and SD DM interactions. The study explores fourteen independent EFT operators, parameterised by Wilson coefficients, to characterise DM interactions with SM particles. Different operators correspond to different interaction mechanisms, and mapping their constraints ensures that no viable scenario is overlooked. We determine the corresponding sensitivity ceiling for each interaction channel as a result of attenuation of the DM flux incident on the detector, affected by DM scattering in the Earth and atmosphere. These ceilings represent the boundaries beyond which meaningful exclusions are no longer possible. 

The fundamental SI and SD interactions, represented by operators $\mathcal{O}_1$ and $\mathcal{O}_4$, define velocity- and momentum-independent scattering. The exclusion limits for these operators compare the sensitivity of QUEST-DMC with existing results.
The upper bound of the projected 90\% confidence level in the SD cross section reaches $6.5 \times 10^{-24}~{\rm cm}^2$ for the readout based on SQUID under the SLP model.
The corresponding SI exclusion limits extend down to $\sim 7.5 \times 10^{-27}\,{\rm cm}^2$ within $0.025-5\,{\rm GeV}/c^2$. These results confirm QUEST-DMC's ability to explore parameter space beyond what has been achieved by existing direct detection experiments.

Beyond the fundamental interactions, additional EFT operators probe other aspects of DM-nucleon interactions, with $\mathcal{O}_8$ and $\mathcal{O}_{11}$ introducing SI velocity and momentum dependence, respectively. QUEST-DMC provides exclusion sensitivity across multiple orders of magnitude.
The highest lower limit on SI interactions is set by $\mathcal{O}_5$, which incorporates velocity- and momentum-dependent terms, while SD interactions are constrained for $\mathcal{O}_7$ and $\mathcal{O}_{10}$. Simultaneous velocity- and momentum-dependent effects, characterised by $\mathcal{O}_3$ and $\mathcal{O}_6$, extend to a larger scale than $\mathcal{O}_5$. Notably, $\mathcal{O}_{15}$ exhibits only a lower bound, as no upper sensitivity ceiling is encountered within the explored cross-section range. In contrast, the remaining operators are constrained within finite exclusion windows, bounded from below by the projected sensitivity and from above by attenuation-induced ceilings.

We systematically express the constraints derived in this work in terms of relativistic DM–nucleon couplings. This is achieved by matching each Lorentz-invariant bilinear interaction to its corresponding set of non-relativistic EFT operators, as outlined in Tables~\ref{tab:DM_SI} to \ref{tab:DM_tensor}. The resulting classification into four interaction categories reflects the organisation of relativistic bilinears according to the spinor structure of the DM and nucleon currents, including scalar, pseudoscalar, axialvector, and tensor terms. 
In particular, tensor-scalar and scalar-tensor spinor structures, which map to combinations involving $\mathcal{O}_1$, $\mathcal{O}_3$, and $\mathcal{O}_5$, produce projected sensitivities stronger than traditional SI or SD scenarios, with coupling limits at $m_\chi = 1\,{\rm GeV}/c^2$ reaching $3.0 \times 10^{-5}$ and $2.8 \times 10^{-5}\,{\rm GeV}^{-2}$, respectively. Below $0.1\,{\rm GeV}/c^2$, scalar–scalar interactions retain the lowest coupling limits, although scalar–tensor remains more sensitive than other structures across the entire mass range.

Overall, we show that QUEST-DMC will cover a large, yet unexplored parameter space of fourteen Wilson coefficients, refining the landscape of DM interactions whilst including attenuation of the DM flux in the earth and the atmosphere. These constraints serve as a crucial input for the theoretical model building, guiding the development of viable DM candidates. Identifying regions where constraints are weak can also reveal experimental blind spots, motivating new detection strategies and guiding the development of viable DM candidates.
\acknowledgments
This work was funded by UKRI EPSRC and STFC (Grants ST/T006773/1, ST/Y004434/1, EP/P024203/1, EP/W015730/1 and EP/W028417/1), as well as the European Union's Horizon 2020 Research and Innovation Programme under Grant Agreement no 824109 (European Microkelvin Platform). S.A. acknowledges financial support from the Jenny and Antti Wihuri Foundation. M.D.T acknowledges financial support from the Royal Academy of Engineering (RF/201819/18/2). J.S. acknowledges support from the UK Research and Innovation Future Leader Fellowship~MR/Y018656/1. A.K. acknowledges support from the UK Research and Innovation Future Leader Fellowship MR/Y019032/1.
\bibliographystyle{JHEP}
\bibliography{EFTBib}
\end{document}